\documentclass[a4paper, 12pt]{article}

\usepackage[english]{babel}
\usepackage{standalone}
\usepackage[T1]{fontenc}
\usepackage[top=2cm,bottom=2cm,left=2cm,right=2cm]{geometry}

\usepackage{amsmath, amssymb, amsthm, amsbsy, amsfonts, natbib, cases}
\usepackage{graphicx, comment, enumitem}
\usepackage[linesnumbered,ruled,vlined]{algorithm2e}
\usepackage[noend]{algpseudocode}
\usepackage{indentfirst, ragged2e} 
\usepackage[colorinlistoftodos]{todonotes}
\usepackage[colorlinks=true, allcolors=blue]{hyperref}
\usepackage[font=small]{caption}
\usepackage[font=small]{caption}
\usepackage{calligra, lscape}
\usepackage{mathrsfs}  
\usepackage{tikz, wrapfig}
\usepackage[compact]{titlesec}
\titlespacing{\section}{0pt}{*0}{*0}
\titlespacing{\subsection}{0pt}{*0}{*0}
\titlespacing{\subsubsection}{0pt}{*0}{*0}

\theoremstyle{definition}

\newtheorem{lem}{Lemma}

\newtheorem{pro}{Proposition}
\newtheorem{rem}{Remark}

\newcommand{\Y}{\mathbf{Y}}

\newcommand{\E}{\mbox{E}}
\newcommand{\V}{\mbox{V}}

\newcommand{\bs}{\boldsymbol}

\newcommand{\1}{\mathbf{1}}

\allowdisplaybreaks 
\usepackage[nodisplayskipstretch]{setspace}

\begin{document}
\title{\Large{\textbf{Continuous-time Markov-switching GARCH Process with Robust and Efficient State Path and Volatility Estimation}}}

\author{}
\author{Yinan Li$^1$, Fang Liu$^1$
\thanks{$^1$Department of Applied and Computational Mathematics and Statistics, University of Notre Dame, Notre Dame, IN 46556, U.S.A. $^*$Email: Fang.Liu.131@nd.edu}}
\date{}

\maketitle
\vspace{-24pt}

\setstretch{1.5}
\begin{abstract}
We propose a continuous-time Markov-switching  generalized autoregressive conditional heteroskedasticity (COMS-GARCH) process for handling irregularly spaced time series (TS) with multiple volatilities states. We employ a Gibbs sampler in the Bayesian framework to estimate the COMS-GARCH model parameters, the latent state path and volatilities. To improve the inferential robustness and computational efficiency for obtaining the maximum a posteriori estimates for the state path and volatilities, we suggest a multi-path sampling scheme and incorporate the Bernoulli noise injection in the computational algorithm. We provide theoretical justifications for the improved stability and robustness with the Bernoulli noise injection through the concept of ensemble learning and the low sensitivity of the objective function to external perturbation in the TS. We apply the  proposed COMS-GARCH process and the computational procedure to simulated TS, a real currency exchange rate TS, and a real blood volume amplitude TS. The empirical results demonstrate that the COMS-GARCH process and the computational procedure are able to predict volatility regimes and volatilities in a TS with satisfactory accuracy.
\newline

\end{abstract}

\setstretch{1.8}
\section{Introduction}\label{sec:intro}
Heteroskedasticity is a common issue in time series (TS) data.  The generalized autoregressive conditional heteroskedasticity (GARCH) model is a popular discrete-time TS model that accommodates heteroskeasticity and estimates the underlying stochastic volatility. 
The GARCH process has been extensively studied from both the theoretical and practical perspectives \citep{IGARCH,EGARCH,GJRGARCH,TGARCH, QGARCH}. 
Depending on the settings and problems, practical TS data can be recorded at irregularly spaced time points, creating a demand for continuous-time models.  \citet{COGARCH1990}  derived the conditions under which the discretized-time GARCH model converges in distribution to a bivariate non-degenerate diffusion process as the length of the discrete time intervals goes to zero.  The fact that the limiting process consisting of two independent Brownian motions (that drives the underlying volatility process and the accumulated TS, respectively) contradicts the GARCH model's intuition that large volatilities are feedback of large innovations.  \cite{corradi2000reconsidering} applied different parameterizations as a function of the discrete time interval to GARCH$(1, 1)$ and obtained both degenerate and non-degenerate diffusion limits. \citet{COGARCH2002} further showed the asymptotic non-equivalence between the GARCH model and the continuous-time bivariate diffusion limit except for the degenerate case in \cite{corradi2000reconsidering}.  
\citet{COGARCH2004} proposed a COntinuous-time GARCH (CO-GARCH) model that replaces the Brownian motions by a single L\'{e}vy process and incorporates the feedback mechanism by modeling the squared innovation as the quadratic variation of the L\'{e}vy process. Regarding the inferences for the CO-GARCH process, there exist several approaches, such as the  quasi-likelihood \citep{bollerslev1994arch}, method of moments (MoM) \citep{COGARCH2004}, pseudo-likelihood \citep{COGARCHMLE2008,marin2015data}, and Markov chain Monte Carlo (MCMC) procedures  \citep{COGARCH2010}.

Though the GARCH process and its variants, including the CO-GARCH process, account for the conditional heteroskedasticity, studies by \citet{lamoureux1990persistence, IGARCH2004, bauwens2014bayesian} and others have shown that volatility predictions by GARCH-type models may fail to capture the true variation in the presence of regime changes in the volatility dynamics. 
To solve this issue, \citet{MSGARCH1996} developed Markov-switching GARCH (MS-GARCH) model that employs a hidden discrete  Markov chain to assign a state  to each time point. This model generates a temporal state path to allow variations in volatility. Regarding the inferences for the MS-GARCH models, since the state path is unobservable whereas the GARCH model parameters are state dependent, regular likelihood-based approaches require summing over exponentially many possible paths and can be computationally unfeasible. Several alternatives exist that can deal with the problem more efficiently. For example, the collapsing procedures \citep{kim1994dynamic, MSGARCH1996, dueker1997markov, klaassen2002improving} based on simplified versions of the MS-GARCH model incorporates recombination mechanisms of the state space are popular approaches.   \citet{MSGARCH2004} developed a new MS-GARCH model that is analytically tractable and allows the derivation of stationarity condition and the process properties.  \citet{MSGARCH2014} employed a Markov Chain Expectation Maximization (MC-EM) approach. \citet{MSGARCH2010} proposed a Bayesian MCMC method but it can be slow in convergence. Recent methods focus on efficient sampling of state paths. \citet{MSGARCH2012} introduced a Viterbi-based technique to sample state paths; \citet{MSCPGARCH2014} proposed a particle MCMC algorithm; \citet{MSGARCH2016} used a  multi-point sampler in combination with the forward filtering backward sampling technique. Both the likelihood-based and the MCMC estimations have been implemented in software, such as the R package \texttt{MSGARCH} \citep{ardia2019markov}, to facilitate the applications of MS-GARCH models in practice.

To the best of our knowledge, there does not exist a MS-GARCH model for irregularly spaced TS nor a CO-GARCH model to handle multiple states. On the other hand, there is a practical need for continuous-time Markov-switching GARCH processes to analyze TS data that exhibit heteroskedasticity and multiple regimes, and are collected in irregularly spaced time points or  on a near-continuous time scale. For example, a heart rate  variability (HRV) TS can have multiple volatilities regimes due to different types of activities or stress levels, and are typically recorded on a  millisecond scale. Seismic waves are another example of TS data that consist of wave types of different magnitudes and are collected with high-frequency (milliseconds). Seismic waves inform our understanding of earth's interior structure and are also used to predict earthquakes. Financial data are often irregularly spaced in time due to weekend and holiday effects and are known to exhibit different volatility states -- changing behaviour drastically from steadily trending to extremely volatile after a major event or news. 

To fill the methodological gap and respond to the practical needs, we propose a COntinuous-time  Markov-switching GARCH process, COMS-GARCH for short. The  COMS-GARCH model employs the L\'{e}vy process to model volatility within each different state and the continuous-time hidden Markov chain to model  switching between states. For inferences about the COMS-GARCH parameters, we propose a Gibbs sampler  in the Bayesian framework. For maximum a posterior (MAP) estimation on the state path and volatilities, we develop a computational  procedure with a multi-path sampling scheme and the Bernoulli noise injection (NI) to accelerate the optimization and improve the robustness of the predicted state path and volatilities. We provide theoretical justifications to  the employment of the Bernoulli NI, which includes ensemble learning of the state path and lowered sensitivity of the objective function to small random external perturbation in the TS. 

The proposed COMS-GARCH process can be used to  analyze multi-state irregularly spaced TS, estimate the historical volatilities, identify an optimal state path, and forecast volatilities and states.  Though there exist simple and model-free approaches such as realized volatility to estimate the historical volatility, these methods cannot systematically identify different volatility states. In addition, for some TS, such as the HRV TS and seismic wave TS, it may not be meaningful to calculate aggregate measures across time points to calculate the realized volatilities. Though the existing MS-GARCH model can also be used for identify different volatility regimens, it cannot analyze irregularly spaced TS. Finally, we expect the estimated state path and  historical volatilities via our proposed COMS-GARCH model to be more robust, largely due to the computational procedure we design specifically for obtaining the MAP estimation of the state path and volatilities.  


In what follows, we introduce the COMS-GARCH process in Sec \ref{sec:comsgarch} and the inferential approach and computational algorithm in Sec \ref{sec:inference}, including the Bernoulli noise injection technique and its the theoretical properties, and forecasting through the COMS-GARCH model. We demonstrate the applications of the COMS-GARCH process  and the proposed computational procedure via the simulation studies in Sec \ref{sec:simulation}, a real currency exchange rate TS and a real blood volume amplitude TS in Sec \ref{sec:case}. The paper concludes in Sec \ref{sec:discussion} with some final remarks.

\section{COMS-GARCH Process}\label{sec:comsgarch}
We propose the COMS-GARCH process   based upon the CO-GARCH(1,1) process \citep{COGARCH2004}.  To the best of our knowledge,  CO-GARCH$(1, 1)$ is the only L\'{e}vy-process driven CO-GARCH model that has analytical solutions for the model parameters from  the stochastic differential equations and is inference-capable in the context of pseudo-likelihood. \citet{COGARCHPQ2006} theoretically analyzed the  CO-GARCH$(p, q)$ model driven by the L\'{e}vy process for general $p$ and $q$ values, but unable to obtain inferences for the model parameters. 

Let $G_t$ for $t\in (0,T)$ denote the observed TS, and  $L$ refer to the innovation, modeled by a L\'{e}vy process. Our proposed COMS-GARCH process  $(G,\sigma^2, S)= (\{G_t\},\{\sigma^2_t\},\{s_t\})$ for  $t>0$ is the solution to the following set of stochastic differential equations
\begin{numcases}{}
dG_t=Y_t=\sigma_t d L_t(s_t)\label{eqn:sde1}\\
d \sigma_t^2=\alpha(s_t) dt-\beta(s_t)\sigma_{t-}^2d t+\lambda(s_t)\sigma_{t-}^2 d [ L, L ]_{t-} \label{eqn:sde2}\\
\Pr(s_t=j|s_{t-}=k)=\eta_{jk}dt+o(dt)\mbox{ for } j\ne k\label{eqn:sde3}\\
\Pr(s_t=k|s_{t-}=k)=\textstyle 1-\sum_{j\neq k}\eta_{jk}dt+o(dt).\label{eqn:sde4}
\end{numcases}
The increment of the L\'{e}vy process $d L_t(s_t)$ in Eqn (\ref{eqn:sde1}) is assumed standardized with mean 0 and variance 1, and  $[L,L]_{t-}$ in Eqn (\ref{eqn:sde2}) is its quadratic variation process.   $\sigma^2_t$ is the underlying volatility process governing the state of $G_t$ at time $t$. Eqns (\ref{eqn:sde3}) and (\ref{eqn:sde4}) represent the hidden continuous-time Markov chain  with $\nu$ discrete states and transition parameters ${\bs\eta}=\{ \eta_{jk}\}$ that model the regime switching in the TS for $j,k\in\{1,\cdots,\nu\}$ (note that $\eta_{jk}$ in Eqns (\ref{eqn:Sn1}) and (\ref{eqn:Sn2}) is not a probability, and the parameter space for $\eta_{jk}$ is $(0,\infty)$ instead of $\in(0,1)$). 

Next, we define a family of discrete-time processes that approximates the above  continuous-time process $(G,\sigma^2, S)$, following the methodological framework in \cite{COGARCHMLE2008}. There are a couple of reasons for obtaining a discretized process. First, real-life observed TS data are often recorded in discrete time, whether irregularly spaced or regardless of how fine the time scale is.  Second, the discretization allows us to take advantage of the well developed inferential approaches  for discrete-time GARCH processes.  We will show the discretized process converge to the COMS-GARCH process.  

The discretization is defined over a finite time interval $[0,T]$ for $T > 0$. Let $0=t_0<t_1<\cdots<t_i<\cdots<t_n=T$ be a deterministic sequence that divides $[0,T]$ into $n$ sub-intervals of lengths $\Delta t_i=t_i-t_{i-1}$ for integers $i=1,\ldots,n$. Let $G_0\!=\!0$ and $\epsilon_i$ be a first-jump approximation of the L\'{e}vy process \citep{COGARCH2004}. A discretized COMS-GARCH process $(G_n,\sigma_n^2, s_n)= (\{G_i\},\{\sigma^2_i\},\{s_i\})$ satisfies
\begin{numcases}{}
G_i-G_{{i-1}}=Y_i=\sigma_{i-1}\sqrt{\Delta t_i} \epsilon_i,\label{eqn:Gn}\\
\sigma_i^2=\alpha(s_i) \Delta t_i +\left(\sigma_{{i-1}}^2+\lambda(s_i)Y_i^2\right)\exp\left(-\beta(s_i)\Delta t_i\right),\label{eqn:sigma2n}\\
\Pr(s_i=j|s_{{i-1}}=k)=1-\exp(-\eta_{jk}\Delta t_i)\mbox{ for } j\neq k,\label{eqn:Sn1}\\
\Pr(s_i=k|s_{{i-1}}=k)=\textstyle\sum_{j\neq k}\exp(-\eta_{jk}\Delta t_i).\label{eqn:Sn2}
\end{numcases} 
Since  $Y_i$ is obtained by differencing the observed $G_i$, it is also observed.  To ensure the positivity of Eqn (\ref{eqn:sigma2n}), we require $\alpha(k)$ and $\lambda(k)$  to be non-negative for all states $k\!=\!1,\ldots,\nu$. To reflect the general belief that dependence between two quantities at two time points diminishes as the time gap increases, we also impose positivity on $\beta(k)$ for all states.  As $n\rightarrow\infty$,  $\Delta t_i\rightarrow0$ and the discretized COMS-GARCH process in Eqns (\ref{eqn:Gn}) to (\ref{eqn:Sn2}) converges in probability to the  COMS-GARCH process defined in Eqns (\ref{eqn:sde1}) to (\ref{eqn:sde4}), as stated in Lemma \ref{lem:conv}. 
\begin{lem}[\textbf{Convergence of discretized COMS-GARCH process}]\label{lem:conv}
Let $(G,\sigma^2, s)$ be the COMS-GARCH process on time interval $[0,T]$, and $(G_n,\sigma^2_n, s_n)$ be its discretized process. 
As $n\rightarrow\infty$, $\Delta t_i\rightarrow 0$ for $i=1,\ldots,n$ and $(G_n,\sigma^2_n, s_n)$ converges in probability to $(G,\sigma^2,s)$ in that the Skorokhod distance $D_S((G_n,\sigma^2_n, S_n),(G,\sigma^2,s)) \overset{p}{\rightarrow} 0\mbox{ as } n\rightarrow\infty$.
\end{lem}
The converges in probability in  Lemma \ref{lem:conv} also implies the convergence of $(G_n,\sigma^2_n, s_n)$ in distribution to $(G,\sigma^2,s)$. Lemma \ref{lem:conv} is an extension of the theorem on the convergence of a discretized CO-GARCH process in  \citet{MLE2008}. The additional complexity in the COMS-GARCH process case is that it accommodates multiple states and has state-dependent GARCH parameters, both of which have no material impact on the discretization of the process and the underlying conditions for the convergence of discretized GARCH process. Therefore,  the theoretical result of the convergence of the discretized COGARCH process n  \citet{MLE2008} can be directly extended to the discretized COMS-GARCH process.  
\begin{rem}
The CO-GARCH model can be regarded as a special case of the COMS-GARCH process with the number of states $\nu=1$. The inferential approaches proposed for COMS-GARCH in Sec \ref{sec:pl} and \ref{sec:bayesian} and some theoretical results in Sec \ref{sec:theory} also apply to CO-GARCH.
\end{rem}

\section{Inferences for COMS-GARCH Process}\label{sec:inference}
The parameters in the COMS-GARCH process include  $\Theta=\{\alpha(k),\beta(k),\lambda(k)\}\;\forall\; k=1,\ldots, \nu$ and transition parameters $\bs\eta$. In addition to $\Theta$ and $\bs\eta$, we are also interested in learning the latent state $s_i$ and volatility $\sigma^2_i$ for $i=1,\ldots,n$ so to better understand an observed TS and to aid prediction of future states and volatilities. 

As presented in Sec \ref{sec:intro}, there exist several approaches for obtaining inferences about the CO-GARCH and MS-GARCH models, respectively, such as the MoM, the maximum likelihood estimation (MLE) based on quasi-likelihood and the pseudo-likelihood, and the Bayesian  MCMC framework for the CO-GARCH model, and the collapsing procedures, MLE via the EM  or MC-EM algorithms, and Bayesian MCMC algorithms for the MS-GARCH model. For inferences about the proposed COMS-GARCH process, which has both the ``MS'' and the ``CO'' components, we propose a Bayesian Gibbs sampler coupled with the pseudo-likelihood as the inferential approach for the COMS-GARCH process. The rationales are as follows.  The MoM for the CO-GARCH model does not work as it cannot deal with irregularly spaced TS. The collapsing procedure introduces an approximation into the MS-GARCH model and a bias in the likelihood \citep{augustyniak2014maximum}. The MLE based on the pseudo-likelihood, pseudo-likelihood, and from the EM algorithms do not provide straightforward solutions on latent states $s_i$ and volatilities $\sigma^2_i$ which are also of inferential interests in addition to $\Theta$ and $\bs\eta$, whereas the Bayesian framework is rather convenient to achieve that goal. 

In what follows, we first present the pseudo-likelihood for the  COMS-GARCH  process in Sec \ref{sec:pl}, then develop a Gibbs sampler for obtaining Bayesian inferences for the model parameters and estimating the latent states and volatilities from the COMS-GARCH process in Sec \ref{sec:bayesian}. We propose a robust and efficient computational procedure for obtaining the MAP estimates for the state path and volatilities in Sec \ref{sec:reSAVE} with multi-path sampling and the Bernoulli NI. A  theoretical analysis of the effects of the Bernoulli NI on learning COMS-GARCH from a TS in Sec \ref{sec:theory}. A foresting procedure on future states and volatilities are provided in Sec \ref{sec:forecasting}.

\subsection{Pseudo-likelihood for COMS-GARCH Process}\label{sec:pl}
Extending the pseudo-likelihood framework in  \citet{MLE2008} for CO-GARCH to COMS-GARCH with multiple states, we have for $i=1,\ldots,n$, that\vspace{-12pt}
\begin{align}
&f(Y_i|Y_1,\ldots,Y_{{i-1}}, s_1,\ldots, s_i)= N(0, \rho_i^2),\mbox{ where}\label{eqn:pseudo}\\
\rho_i^2&\!=\!\left(\! \sigma^2_{{i-1}}\!-\!\frac{\alpha(s_i)}{\beta(s_i)-\lambda(s_i)} \!\right)\!\left( \frac{\exp((\beta(s_i)\!-\!\lambda(s_i))\Delta t_i)\!-\!1}{\beta(s_i)-\lambda(s_i)} \right)\!+\!\frac{\alpha(s_i)\Delta t_i}{\beta(s_i)-\lambda(s_i)}\label{eqn:exact}\\
&\approx \sigma^2_{{i-1}}\Delta t_i= \alpha(s_{{i-1}})\Delta t_{i-1} \Delta t_i+\Delta t_i\left(\sigma_{{i-2}}^2+\lambda(s_{{i-1}})Y_{{i-1}}^2\right)\exp\left(-\beta(s_{{i-1}})\Delta t_{i-1})\right).\label{eqn:approx}\vspace{-12pt}
\end{align}
Eqn (\ref{eqn:approx})  is obtained by taking the first-order Taylor expansion of Eqn (\ref{eqn:exact}) around $\Delta t_i=0$ and  substituting $\sigma_{i-1}^2$ in Eqn (\ref{eqn:sigma2n}). 
Eqns (\ref{eqn:pseudo}) to (\ref{eqn:approx}) suggest that \vspace{-12pt}
\begin{equation}\label{eqn:EY2}
\E(Y^2_i|Y_1,\ldots,Y_{{i-1}}, s_1,\ldots, s_i)=\V(Y_i|Y_1,\ldots,Y_{{i-1}}, s_1,\ldots, s_i)=\rho_i^2\approx\sigma_{i-1}^2\Delta t_i, \vspace{-12pt}
\end{equation}
which will be useful for foresting future volatilities as demonstrated in Sec. \ref{sec:bayesian}.

\subsection{Gibbs Sampler for Bayesian Inferences about Model Parameters}\label{sec:bayesian}
If the goal of fitting the COMS-GARCH process to a TS is to obtain inferences for $\Theta$ and $\bs\eta$ given the pseudo-likelihood, the EM and MC-EM algorithms would make good choice to obtain Pseudo-MLE for $\Theta$ and $\bs\eta$, integrating out the latent states (we provide the EM and MC-EM algorithms in the Supplementary Materials for interested readers). As mentioned, our interest is also in estimating and forecasting states and volatilities, methods that rely on integrating out the latent states do not work well, whereas the Bayesian framework provides a convenient and straightforward approach.  We propose a Gibbs sampler to obtain Bayesian inferences for the COMS-GARCH process.

Define $\bs{\Delta t}\!=\!(\Delta t_1,\ldots,\Delta t_n),\Y\!=\!(Y_1,\ldots,Y_n), S
\!=\!(s_1,\ldots,S_n)$, and $\mathcal{S}$ is the set of all possible state paths. Denote the priors for $\Theta$ and $\bs\eta$ by $\pi(\Theta,\bs\eta)$ and  assume $\pi(\Theta,\bs\eta)\!=\!\pi(\Theta)\pi(\bs\eta)$.  The conditional posterior distributions of $\Theta,\bs{\eta}$, and the states are respectively \vspace{-12pt}
\begin{align}
&f(\Theta|\bs\eta, \Y,\Delta \mathbf{t},S)\propto \pi(\Theta)L(\Theta,\bs\eta|\Y,S)=\textstyle  \pi(\Theta)\prod_{i=1}^{n}\rho_i^{-1}\exp\left( -Y_i^2/(2\rho_i^2) \right),\label{eqn:fTheta}\\
&f(\eta_{1k},\ldots,\eta_{\nu k}|\Theta,\Y,\Delta\mathbf{t},S)
=\textstyle f(\eta_{1k},\ldots,\eta_{\nu k}|S,\Delta\bs{t})\mbox{ for } k=1,\ldots,\nu, \notag\\
&\propto\pi(\eta_{1k},\ldots,\eta_{\nu k})\!\!\!\!\!\!\!\!
\prod_{s_{i+1}=k,s_i=k}^{n-1}\!\!\!\left(\!2\!-\!\nu\!+\!\sum_{j\ne \nu}\exp(-\eta_{jk}\Delta t_{i+1})\right)\prod_{j\ne k}
\prod_{s_{i+1}=j,s_i=k}^{n-1}\!\!\!\!\!\!(1\!-\!\exp(-\eta_{jk}\Delta t_{i+1})),\label{eqn:feta}\\
&f({s_i}|S_{-i},\Theta,\bs\eta, \Y,\mathbf{\Delta t})
\propto\textstyle \xi_{s_i,s_{{i-1}}}\xi_{s_{i+1},s_i}
\prod_{t=i}^{n}\rho_t^{-1}\exp\!\left(\!-
Y_t^2/(2\rho_t^2) \right),\label{eqn:fs}
\end{align}
where $\rho_i^2$ in Eqn (\ref{eqn:fTheta}) is a  function of $\Theta$ (see Eqn (\ref{eqn:approx})), $\sum_{j=1}^{\nu}\eta_{jk}=1$ in Eqn (\ref{eqn:feta}), and
$$\xi_{s_i,s_{i-1}}\!=\!\begin{cases}\begin{aligned}
\textstyle 2-\nu+\sum_{k\ne  s_{i-1}}\exp(-\eta_{k,s_{i-1}}\Delta t_i) & \mbox{ when }  s_i=s_{i-1}\\
1-\exp(-\eta_{s_i,s_{i-1}}\Delta t_i) & \mbox{ when }s_i\neq s_{i-1}\\
\end{aligned}
\end{cases}\!\!\!\!\!;
\mbox{ similarly for $\xi_{s_{i+1},s_i}$. When there }$$
are two states ($\nu=2$),  Eqns (\ref{eqn:feta}) and (\ref{eqn:fs}) can be simplified to\vspace{-9pt}
\begin{align}
f(\eta_{21}|S,\bs{Y},\Delta\bs{t})\propto&\textstyle \pi(\eta_{21})\prod_{s_{{i+1}}=1,s_i=1 }^{n-1}\exp(-\eta_{21}\Delta i) \prod_{s_{i+1}=2,s_i=1 }^{n-1}(1-\exp(-\eta_{21}\Delta t_{i+1}))\label{eqn:transrate1}\\
f(\eta_{12}|S,\bs{Y},\Delta\bs{t})\propto&\textstyle  \pi(\eta_{12})\prod_{s_{{i+1}}=2,s_i=2 }^{n-1}\exp(-\eta_{12}\Delta i) \prod_{s_{{i+1}}=1,s_i=2 }^{n-1}(1-\exp(-\eta_{12}\Delta t_{i+1}))\label{eqn:transrate2}\\
f(s_i|S_{-i},\Theta,\bs\eta,\!, \Delta\mathbf{t})&\textstyle\propto\! \xi_{1,s_{i-1}}^{2-s_i}\! \xi_{2,s_{i-1}}^{s_i-1}\!\xi_{1,s_i}^{2-s_{i+1}} \!\xi_{2,s_i}^{s_{i+1}-1}\! \prod_{t=i}^{n}\rho_t^{-1}\exp\!\left(\!-
\frac{Y_t^2}{2\rho_t^2} \!\right),\vspace{-9pt}
\end{align}
where $\xi_{1,s_{i-1}}\!=\!\exp(-\eta_{21}\Delta t_i)$ if $s_{i-1}\!=\!1$, and  $1-\exp(-\eta_{21}\Delta t_i)$ if $s_{i-1}\!=\!2$;  $\xi_{1,s_i} \!=\!\exp(-\eta_{21}\Delta t_{i+1})$ if $s_i=1$, and $1-\exp(-\eta_{21}\Delta t_{i+1})$ if $s_i=2$.

The Gibbs sampler draws samples on $\Theta,\bs{\eta}$ and $s_i$ for $i=1,\ldots,n$ alternatively from Eqns (\ref{eqn:fTheta}), (\ref{eqn:feta}), and (\ref{eqn:fs}). Upon convergence, after burning and thinning, we will have multiple, say $M$,  sets of posterior samples of $\Theta,\bs{\eta}$, based on which their posterior inferences can be obtained. We will also have $M$ sets of samples on state $s_i$ and can calculate the posterior  volatility $\sigma_i^2$ at each time point via Eqn (\ref{eqn:sigma2n}). Connecting the states across the $n$ times points from each set of the state posterior samples leads to a state path.  
Due to the large sample space (totally  $\nu^n$ possible paths), it is difficult to identity the MAP  estimate for the state path out of the $M$ paths with acceptable accuracy unless $M\!>\!\!>\!\mu^n$ and a significant portion of paths have close-to-0  posterior probabilities with a few paths having significantly higher probabilities compared to the rest. To solve this issue, we design a new computational algorithm (reSAVE) as detailed next.

\subsection{MAP Estimation for State Path and Volatility} \label{sec:reSAVE}
To deal with the computational challenge in obtaining the MAP estimates for state path via the Gibbs sampler in Sec \ref{sec:bayesian}, we propose an inferentially \textbf{R}obust and computationally \textbf{E}fficient iterative procedure for \textbf{S}tate path \textbf{A}nd \textbf{V}olatility \textbf{E}stimation (reSAVE). 
The inferential robustness for the MAP estimates of the reSAVE procedure is brought by the Bernoulli NI  implemented in each iteration of procedure, leading to both ensemble learning and improved stability of the object function from which the MAP estimates are obtained (see Sec \ref{sec:theory} for details). The computational efficiency of the reSAVE can be attributed to a couple of factors: the Bernoulli NI that generates a sub-TS (smaller data size) in each iteration, of sampling of a small set of state $m$ to calculate MAP estimates for the state path, and the employment of a maximization-maximization scheme to obtain the MAP estimates of the parameters $\Theta$ and $\bs\eta$ and for the state path in each iteration.  The steps of the reSAVE procedure  are listed in Algorithm \ref{alg:reSAVE}.  Remarks \ref{rem:Nm} and \ref{rem:MAP} provide further remarks and explanations on some steps of the procedure. 
\begin{algorithm}[!htb]
\caption{The reSAVE Optimization Procedure}\label{alg:reSAVE}
\SetAlgoLined
\SetKwInOut{Input}{input}
\SetKwInOut{Output}{output}
\Input{Observed ata $(\Y,\Delta\mathbf{t})$, initial values $\Theta^{(0)},\bs\eta^{(0)},S^{(0)}=\left(s^{(0)}_1,\ldots,s^{(0)}_n\right)$, number of iterations $N$, number of sampled state paths $m$, and ensemble size $b$ (optional)  (Remark \ref{rem:Nm}).}
\Output{MAP estimates $\hat{S}_{\text{MAP}}, \hat{\bs\sigma}_{\text{MAP}}^2\!=\!(\hat\sigma_1^2,\ldots,\hat\sigma_n^2), \hat\Theta_{\text{MAP}}, \hat{\bs\eta}_{\text{MAP}}$.}
\For{$l= 1$ to $N$}{ 
Apply the Bernoulli noise injection in Algorithm \ref{alg:dropout} to obtain a sub-TS $(\tilde{\Y}^{(l)},\Delta{\tilde{\mathbf{t}}}^{(l)})$ of length $\tilde{n}^{(l)}$. Denote  by $\mathcal{T}^{(l)}$ the set of the original time points retained in the sub-TS\;
Calculate $\hat\Theta^{(l)}_{\text{MAP}}=\arg\max\limits_{\Theta}f(\Theta|S^{(l-1)},\tilde{\Y}^{(l)},\Delta{\tilde{\mathbf{t}}}^{(l)})$ and $\mbox{\hspace{48pt}}\hat{\bs{\eta}}^{(l)}_{\text{MAP}}=\arg\max\limits_{\bs\eta}f(\bs\eta|S^{(l-1)},\tilde{\Y}^{(l)},\Delta{\tilde{\mathbf{t}}}^{(l)})$  (Remark \ref{rem:MAP})\;
\For{$j = 1$ to $m$}{
\For{$i = 2$ to  $n-1$}{
if $i\!\in\!\mathcal{T}^{(l)}$, then sample $s^{(j)}_i$  given  $\Theta^{(l)}_{\text{MAP}}$, $\bs\eta^{(l)}_{\text{MAP}}$, $\tilde{\Y}^{(l)},\Delta{\tilde{\mathbf{t}}}^{(l)}, S_{-i}^{*(l-1)}$ per Eqn (\ref{eqn:fs})\;}
Let $\tilde{S}^{(j)}=\left(s^{(j)}_1,s^{(j)}_2,\ldots, s^{(j)}_{\tilde{n}^{(l)}}\right)$.} 
Let $\tilde{\mathcal{S}}\!=\!\!\left(\!\tilde{S}^{(1)},\ldots,\tilde{S}^{(m)}\!\right)$; solve $\tilde{S}^{(l)}\!=\!\arg\max\limits_{S\in \tilde{\mathcal{S}}}f(S| {\Theta}^{(l)}_{\text{MAP}},{\bs\eta}^{(l)}_{\text{MAP}},\tilde{\Y}^{(l)}\!,\Delta{\tilde{\mathbf{t}}}^{(l)})$ (Remark \ref{rem:MAP})\;
Let $S^{(l)}=\{\tilde s_i^{(l)}\}_{i\in\mathcal{T}^{(l)}} \bigcup\{\tilde s_i^{(l-1)}\}_{i\notin\mathcal{T}^{(l)}}$\; 
}
Calculate MAP estimate for volatility $\{\sigma_i^2\}_{i=1,\ldots, n}$  given  $S_{\text{MAP}}\!=\!S^{(N)}, \hat{\Theta}_{\text{MAP}}\!=\!\hat{\Theta}^{(N)}_{\text{MAP}}$, $\hat{\bs\eta}_{\text{MAP}}\!=\!\hat{\bs\eta}^{(N)}_{\text{MAP}}$ via Eqn (\ref{eqn:sigma2n}). 
\end{algorithm}
\vspace{-12pt}
\begin{rem}\label{rem:Nm}
Regarding the number of iterations $N$, either $N$ can be prespecified or a convergence criterion can be used, such as the $l_1$ distances on the MAP estimates (e.g., $\{|\hat{\bs\eta}^{(l+1)}-\hat{\bs\eta}^{(l)}|, |\hat{\Theta}^{(l+1)}-\hat{\Theta}^{(l)}, |S^{(l+1)}-S^{(l)}|\}$) or the objective functions between two consecutive iterations. If the distances are below a certain threshold, then the algorithm converges and can stop.
Regarding the number of sampled state paths $m>1$,  $m$ too small will not lead to stable MAP estimates;  $m$ too large would increase the computational costs. In the simulation and case studies in Secs \ref{sec:simulation} and \ref{sec:case}, we used $m=6$, which is deemed good enough. In general, we suspect the choice of $m$ relates to $n$ -- the number of time points in the observed TS; the larger $n$ is, the larger $m$ is. The ensemble size $b$ refers to the number of observations following a given time point $i$ that are used to update the conditional posterior distribution of $s_i$. The specification of $b$ is mainly for computational efficiency consideration and is optional (see Proposition \ref{prop:ensemble}).
\end{rem}
\begin{rem}\label{rem:MAP} 
The MAP estimates of $\Theta$ and $\bs\eta$ can be determined either through direct optimization of their respective conditional posterior distributions, or via MC approaches based on samples of $\Theta$ and $\bs\eta$ from their conditional posterior distributions.  
The MAP estimate of the state path in each iteration is defined as the path, out of the sampled $m$ paths, that maximizes the conditional posterior distribution of $S$, which is proportional to $\prod_{i=1}^n\rho_i^{-1}\exp(-y_i^2/(2\rho_i^2))\eta_{i,i-1}$ \citep{MSGARCH2010}, given the latest MAP estimates $\Theta$ and $\bs{\eta}$ .
\end{rem}

In Algorithm \ref{alg:reSAVE}, if we set $m$ at 1 and remove line 10 for identifying the MAP estimate for $S$, and replace the calculation of conditional MAP estimates $\hat{\Theta}_{\mbox{\tiny{MAP}}}$ and $\hat{\bs\eta}_{\mbox{\tiny{MAP}}}$ in line 3 by random sampling $\Theta$ and $\bs\eta$  from their respective conditional posterior distributions, then it basically is the Gibbs sampler presented in Sec \ref{sec:bayesian} for obtaining the Bayesian inferences about $\Theta$ and $\bs\eta$, with one important difference -- the usage of Bernoulli NI to generate a sub-TS in every iteration. Though the Bernoulli NI is designed more for achieving ensemble learning and improving the stability of the objective functions for the state path optimization (see Sec \ref{sec:theory}), we expect its usage also helps in making inferences on $\Theta$ and $\bs\eta$ more robust, especially if $\nu$ is relatively large.

A key step in Algorithm \ref{alg:reSAVE} is the generation of sub-TS via the Bernoulli NI. The rationale for creating sub-TS when estimating the state path and volatilities is that the estimation can be sensitive to the TS data for the COMS-GARCH process. We employ the Bernoulli NI to  create an ensemble of sub-TS' of  considerable diversity among the ensemble members across iterations so to reduce the sensitivity (see Sec \ref{sec:theory} for more details). Algorithm \ref{alg:dropout} lists the steps of the Bernoulli NI. Specifically, in each iteration of Algorithm \ref{alg:reSAVE}, we apply the Bernoulli NI to the observed TS $\mathbf{G}=\{G_i\}_{i=0}^n$ to obtain a sub-TS $\tilde{\mathbf{G}}$. Denote the sequence of time gaps in $\tilde{\mathbf{G}}$  by $\Delta \tilde{\mathbf{t}}$. With sub-TS $\{\tilde{\mathbf{G}},\Delta \tilde{\mathbf{t}}\}$, we only need to sample and update the states $\tilde{S}$ at the retained time points in each iteration, and the states of the dropped time points are kept at the values from the previous iteration, saving costs computationally. 
\begin{algorithm}[!htb]
\caption{Bernoulli Noise Injection}\label{alg:dropout}
\SetAlgoLined
\SetKwInOut{Input}{input}
\SetKwInOut{Output}{output}
\Input{Original TS $\mathbf{G}$; Bernoulli NI rate $p$ specified by users or chosen by cross-validation (see Algorithm \ref{alg:cv}).}
\Output{sub-TS $(\tilde{\Y},\Delta{\tilde{\mathbf{t}}},\tilde{n})$.}
Draw $e_i$ independently from Bern$(1-p)$ for $i=2,\ldots,n-1$. Set $e_0=e_1=e_{n}=1$\;
Let $\tilde{\mathbf{G}}=\{\mathbf{G}: \mathbf{G}\cdot \mathbf{e}\neq 0\}$, where  $\mathbf{e}=\{e_i\}_{i=0}^n$,  $\tilde{n}=\sum_{i=1}^{n}e_i$\;
Obtain $\tilde{\Y} =\{\tilde{Y} _1,\ldots,\tilde{Y} _{\tilde{n} }\}=\mbox{diff}(\tilde{\mathbf{G}} )$\;
Let $\Delta \tilde{\mathbf{t}}=\Delta {\mathbf{t}}$. For $0 \leq i\leq n-1$, re-set
$\begin{cases}
\Delta \tilde{t}_{i+1} =\Delta i +\Delta\tilde{t}_{i+1} \mbox{ and }
\Delta i =0 \mbox{ if } e_i=0 \\
\Delta \tilde{t}_{i+1} =\Delta \tilde{t}_{i+1} \mbox{ if } e_i=1\\
\end{cases}$\;
Let $\Delta \tilde{\mathbf{t}} =\{\Delta \tilde{\mathbf{t}}: \Delta \tilde{\mathbf{t}}\neq 0\}$\;
\end{algorithm}
\vspace{-12pt}
\begin{rem}\label{rem:NI}
The differenced $\tilde{\Y}$ in a sub-TS after the Bernoulli NI is a summation of a sequence of differenced $\Y$ formed with the dropped observations in the original TS.
\end{rem}
\vspace{-12pt}
Remark \ref{rem:NI} is a simple but interesting fact.  For example, if $G_{i+1}$ gets dropped from the sequence of $\ldots,G_i, G_{i+1}, G_{i+2},\ldots$, then $\tilde{Y}_{i'}= G_{i+2}-G_{i}= (G_{i+2}-G_{i+1}) + (G_{i+1}-G_{i})= Y_{i+2}+Y_{i+1}$; say $r$ observations are dropped between $G_i$ and $G_{i+r+1}$, then $\tilde{Y}_{i'}= G_{i+r+1}-G_{i}= (G_{i+r+1}-G_{i+r}) + (G_{i+r}-G_{i+r-1}) +\cdots + (G_{i+1}-G_{i})= Y_{i+r+1}+\cdots+Y_{i+1}$. This fact is used in the proof of Proposition \ref{prop:perturb1} in Sec \ref{sec:theory}. Since the NI rate $p$ is usually small and the times points are dropped from the original TS randomly, with the fine time scale on which the TS is collected, the COMS-GARCH process can ``digests'' these ``missing'' time points effortlessly, without needing an ad-hoc approach to handle these dropped time points. The full conditional distributions of $\Theta,\bs{\eta}$, and states $\{s_i\}$ given the sub-TS in each iteration are given in Eqns (\ref{eqn:fTheta}) and (\ref{eqn:fs}) by replacing the original TS $(\Y,\Delta{\mathbf{t}})$ with the sub-TS $(\tilde{\Y},\Delta{\tilde{\mathbf{t}}})$. 

The Bernoulli NI for COMS-GARCH is inspired by the dropout technique for regularizing neural networks (NNs) \citep{dropout14}, which injects Bernoulli noises to the NN structure (hidden and input nodes), leading to the $l_2$ regularization on model parameters.  The Bernoulli NI we propose here is different procedurally in that it is applied to the observed data and drops randomly selected time points in the original TS in each iteration, rather than generating sub-models; in other word,  the  COMS-GARCH  model  remains the same throughout. The benefits of applying the Bernoulli NI in Algorithm \ref{alg:reSAVE} include reduced computational cost and its connection with ensemble learning and inferential stability and robustness. The Bernoulli NI also bears some similarity to bagging \citep{bagging}, a well-known ensemble learning algorithm, but also differs from the latter in two aspects. First,  the Bernoulli NI leads to a random sub-TS (without replacement) of the original TS in each iteration of Algorithm \ref{alg:reSAVE} whereas bagging often generates a bootstrapped sample set with replacement that is of the same size as the training data. Second, bagging often generates multiple sets of samples, trains a model on each set in parallel, and then ensembles them into a meta-model, whereas the  ensemble learning  brought by the Bernoulli NI to the MAP estimation for COMS-GARCH is implicit,  iterative, and realized sequentially.  Finally, the Bernoulli NI we propose is not the same as the down-sampling technique  in TS signal processing \citep{downsampling}. Signal processing aims at extracting useful features from collected signals, where down-sampling is used there for data reduction, compression, memory conservation, among others. 

To choose Bernoulli NI  $p$, we can apply a $k$-fold cross-validation (CV) procedure as listed in Algorithm \ref{alg:cv}. 
\begin{algorithm}[!htb]
\caption{$k$-fold CV for choosing Bernoulli NI rate ${p}$}\label{alg:cv}
\SetAlgoLined
\SetKwInOut{Input}{input}
\SetKwInOut{Output}{output}
Generate $k$ non-overlapping sub-TS' $\Y_{\mbox{cv},1},\ldots,\!\Y_{\mbox{cv},k}$ of the original TS; $\bigcup_{k'=1}^k\!\Y_{\mbox{cv},k'}\!=\!\Y$\;
Specify a grid of Bernoulli NI rates $\mathbf{p}$ of length $J$\;
\For{$j=1,\ldots, J$}{
\For{$k'=1,\ldots, k$}{
Set $\Y_{\mbox{cv},k'}$ as the validation set,  the rest are combined to be the training set $\mathcal{Z}_{k'}$\;
Apply Algorithm \ref{alg:reSAVE} to   $\mathcal{Z}_{k'}$ with Bernoulli NI rate $p_j$ to obtain the MAP estimates on the COMS-GARCH parameters, the volatility predictions, and the MAP across the time points in  $\mathcal{Z}_{k'}$\;
Denote the time points in $\Y_{\mbox{cv},k'}$ by $\mathcal{T}_{k'}$\;
\For{$i \in \mathcal{T}_{k'}$}{
Locate the two time points $t_{i_1}$ and $t_{i_2}$ from $\mathcal{Z}_{k'}$ that are closest to $t_i$ that satisfy $t_{i_1}<t_i$ and $t_i<t_{i_2}$.  Set $s_i=s_{i'}$, where $i'\!=\!\arg\min_{i'\in\{i_1,i_2\}}|t_i\!-\!t_{i'}|$.\; 
Predict $\hat{Y}_i^2$ via Eqn (\ref{eqn:EY2}) in the trained COMS-GARCH model, where $\hat{\sigma}^2_{i-1}$ can be calculated directly if $t_{i-1}\in\mathcal{Z}_{k'}$, or solved via Eqn (\ref{eqn:sigma2n}) given $\hat{Y}_{i-1}^2$  if $t_{i-1}\in\Y_{\mbox{cv},k'}$ (Remark \ref{rem:pred}).}
Calculate the mean squared error (MSE) $l_{j,k'}$ between the observed $\Y^2$ and predicted $\hat{\Y}^2$ in the validation set.
}
Calculate $\bar{l}_j\!\!=\!\!\sum_{k'=1}^k\!l_{j,k'}/(k\!-\!1)$ and its standard error se$_j\!\!=\!\!\sqrt{\!\sum_{k'=1}^k\!(l_{j,k'}\!-\!\bar{l}_j)^2/(k(k\!-\!1))}$.
}
Let $j^*\!=\!\arg\min_j \bar{l}_j$ and $j_{1se}^{*}\!$ be the first index in $\mathbf{p}$ that satisfies  $\bar{l}_{j_{1se}^{*}}\!\ge\!\bar{l}_{j^{*}}\!+\!\mbox{se}_{j^*}\!$ (Remark \ref{rem:cv})\;
Set the Bernoulli NI rate at $p_{j_{1se}^{*}}$.
\end{algorithm} \vspace{-12pt}
\begin{rem}\label{rem:pred}
We predict $Y_i^2\approx \hat{\sigma}_{i-1}^2\Delta t_i$ via Eqn (\ref{eqn:EY2}), where $\hat{\sigma}_{i-1}^2$ is the volatility at the immediately preceding time point that belongs to the training set and can be calculated from Eqn (\ref{eqn:sigma2n}) given $s_{i-1}$ and $Y^2_{i-1}$. Once $\hat{Y}^2_i$ is obtained, $\hat{\sigma}^2_{i}$ can be back-calculated from Eqn (\ref{eqn:sigma2n}) given $\hat{Y}^2_i$ and the interpolated state $s_i$.
\end{rem}
\begin{rem}\label{rem:cv}
We apply the one-standard-error rule when choosing the Bernoulli NI rate $p$ instead of using the one  minimizes the CV error due to two reasons. First, the one-standard-error rule is commonly used in selecting tuning parameters through CV in statistical machine learning to further mitigate over-fitting  and improve  generalization of the trained model. Second, there are always concerns on dependency between  training and validating sets no matter what procedure is used for partitioning a TS into training and validating sets when developing CV procedures for TS data. While some partitioning methods might lead to less dependency than  others, this is often achieved by throw away some data points \citep{hjorth1982model, PCV, chu1991comparison}. We conjecture that the application of the one-standard-error rule (or something even  harsher) helps alleviate the dependency concerns and leads to  more generalizable parameter estimation and more robust prediction.
\end{rem}

\vspace{-12pt}
\subsection{Theoretical Analysis on Inferential Benefits of Bernoulli NI}\label{sec:theory}
We state briefly in Sec \ref{sec:reSAVE} that the Bernoulli NI aims at improving the efficiency and robustness of the MAP estimates of the state path and thus the volatilities. In this section,  we investigate  theoretically the inferential benefits of the Bernoulli NI in two aspects. First, we show that the Bernoulli NI assists in the MAP estimation in an ensemble learning fashion; second, we establish that Bernoulli NI stabilizes the objective function in the presence of random external perturbation in the original TS.

\subsubsection{Ensemble Learning of State Path}\label{sec:ensemble}
Ensemble learning aims for better prediction by assembling or combining a diverse group of learned models given a training data set \citep{ensemble99, ensemble96, ensemble03, ensemble06}. Many  machine learning techniques are built on  ensemble learning such as boosting and bagging.  We show that, through the iterative Algorithm \ref{alg:reSAVE},  the Bernoulli NI leads to sequential and implicit ensemble learning of the parameters and states for the COMS-GARCH model.  The formal results are presented in Proposition \ref{prop:ensemble}. 
\vspace{-9pt} 
\begin{pro} [\textbf{ensemble learning of state path with Bernoulli NI}] \label{prop:ensemble} 
Assume that for $\forall\; \epsilon>0$, $\exists b\in N+$ such that 
\begin{align}\label{eqn:ass}
&\!\!\left\vert\frac
{\prod_{j=i}^nf(Y_j|\Y_{j-1}, s_1,\ldots,s_i\!=\!k_1,\ldots,s_j)}
{\prod_{j=i}^nf(Y_j|\Y_{j-1}, s_1,\ldots,s_i\!=\!k_2,\ldots,s_j)}\!-\!
\frac
{\prod_{j=i}^{i+b}f(Y_j|\Y_{j-1}, s_1,\ldots,s_i\!=\!k_1,\ldots,s_j)}
{\prod_{j=i}^{i+b}f(Y_j|\Y_{j-1}, s_1,\ldots,s_i\!=\!k_2,\ldots,s_j)}\right\vert<\epsilon
\end{align} 
$\forall k_1\neq k_2 \in\{1,\ldots,\nu\}, i\leq n-b, \Y_{j-1}=(Y_1,\ldots,Y_{j-1})$. There exist $C_{k-1}^{b-1}$ ways to yield a set of b observations from a sequence of $k\in[b,n-i]$ consecutive observations. Denote the ensemble of the resultant $C_{k-1}^{b-1}$ sub-TS' by $\tilde{\mathcal{Y}}$. Given a Bernoulli NI rate $p$, the conditional posterior distribution of $s_i$ given the ensemble $\tilde{\mathcal{Y}}$ is\vspace{-9pt} 
\begin{equation}\label{eqn:ensemble}
\textstyle \sum_{k=b}^{n-i} \left(p^{k-b} (1-p)^{b-1} \sum_{\tilde{\Y}\in\tilde{\mathcal{Y}}}f({\tilde{s}_{i } }|{\tilde{S}_{-i } },\Theta,\bs\eta, \tilde{\Y})\right).\vspace{-5pt} 
\end{equation}
\end{pro}
The proof of Proposition \ref{prop:ensemble} is straightforward.  Eqn (\ref{eqn:sigma2n}) implies that the conditional distribution of $Y_j$ depends only on its variance since its mean is fixed at 0.  Eqn (\ref{eqn:pseudo}) suggests that the impact of state $s_i$ on $\sigma^2_{j-1}$ (and thus $\rho_j^2$) decreases as $i$ departs from $j$ given the recursive formula on $\sigma^2$.  Taken together, it implies that the state at time $t_i$ has minimal effect on the distribution  of $Y_j$ at a future time point $t_j$ once the distance  $t_j-t_i$ surpasses a certain threshold, which we use $b$ to denote. Mathematically, it means the ratio between $\prod_{j=i+b+1}^nf(Y_j|\Y_{j-1},s_1,\ldots,s_i\!=\!k_1,\ldots,s_j)$ and $\prod_{j=i+b+1}^nf(Y_j|\Y_{j-1},s_1,\ldots,s_i\!=\!k_2,\ldots,s_j)$ is arbitrarily close to 1, or 
$$\frac
{\prod_{j=i}^{i+b}f(Y_j|\Y_{j-1},s_1,\ldots,s_i\!=\!k_1,\ldots,s_j)}
{\prod_{j=i}^{i+b}f(Y_j|\Y_{j-1},s_1,\ldots,s_i\!=\!k_2,\ldots,s_j)}
\left\vert\frac
{\prod_{j=i+b+1}^nf(Y_j|\Y_{j-1},s_1,\ldots,s_i\!=\!k_1,\ldots,s_j)}
{\prod_{j=i+b+1}^nf(Y_j|\Y_{j-1},s_1,\ldots,s_i\!=\!k_2,\ldots,s_j)}\!-\!1 \right\vert<\epsilon $$
for any $\epsilon>0$, leading to Eqn (\ref{eqn:ass}). $k$ given $b$ and $p$ follows a negative binomial distribution, leading directly to Eqn (\ref{eqn:ensemble}). 

Taken together with the posterior distribution of $s_i$ in Eqn (\ref{eqn:fs}), Eqn (\ref{eqn:ass}) implies the posterior distribution of $s_i$ can be almost surely determined by the $b$ observations in TS $\Y$ that immediately follow $t_i$; that is,
$f(s_i|S_{-i},\Theta,\bs\eta,\Y) = f(s_i|S_{-i},\Theta,\bs\eta,Y_i,Y_{i+1},\ldots, Y_{i+b})$. This narrow focus on just $b$ observations is  undesirable from an inferential perspective especially when $b$ is small because the inference about the state path can be unstable and highly sensitive to even insignificant fluctuation in the TS. The Bernoulli NI helps mitigate this concern by diversifying the set of the $b$ observations.Specifically, after the Bernoulli NI, the posterior probability of $s_i$ in Algorithm \ref{alg:reSAVE} is a weighted average of the posterior distributions over multiple sets of $b$ observations with different compositions across iterations, as suggested by Eqn (\ref{eqn:ensemble}), leading to more robust state estimation. 

\begin{figure}[!htb]
\begin{minipage}{1\textwidth}
\centering
\includegraphics[width=0.325\linewidth, height=0.32\linewidth, trim=0pt 0pt 1cm 0pt, clip]{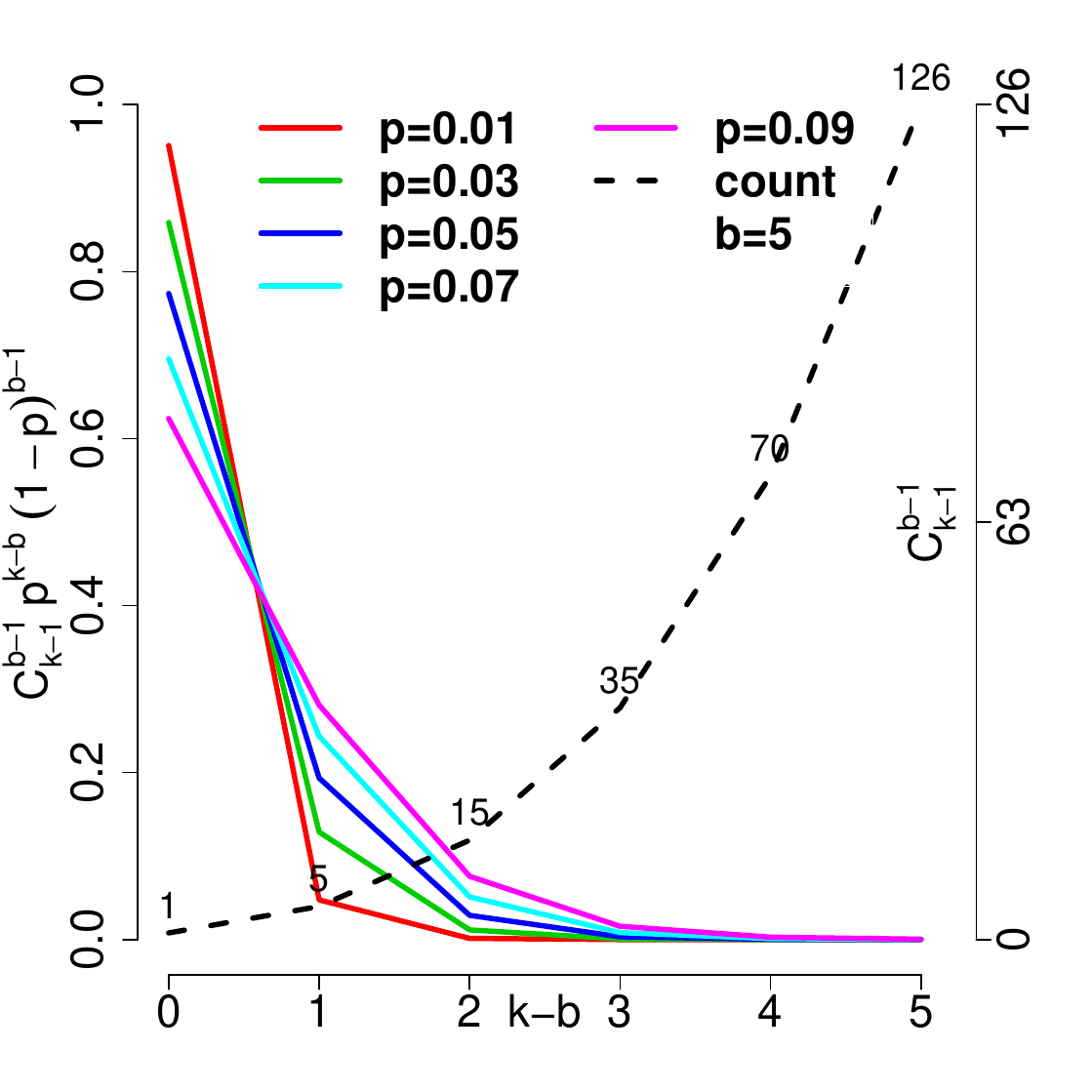}
\includegraphics[width=0.325\linewidth, height=0.32\linewidth, trim=0pt 0pt 1cm 0pt, clip]{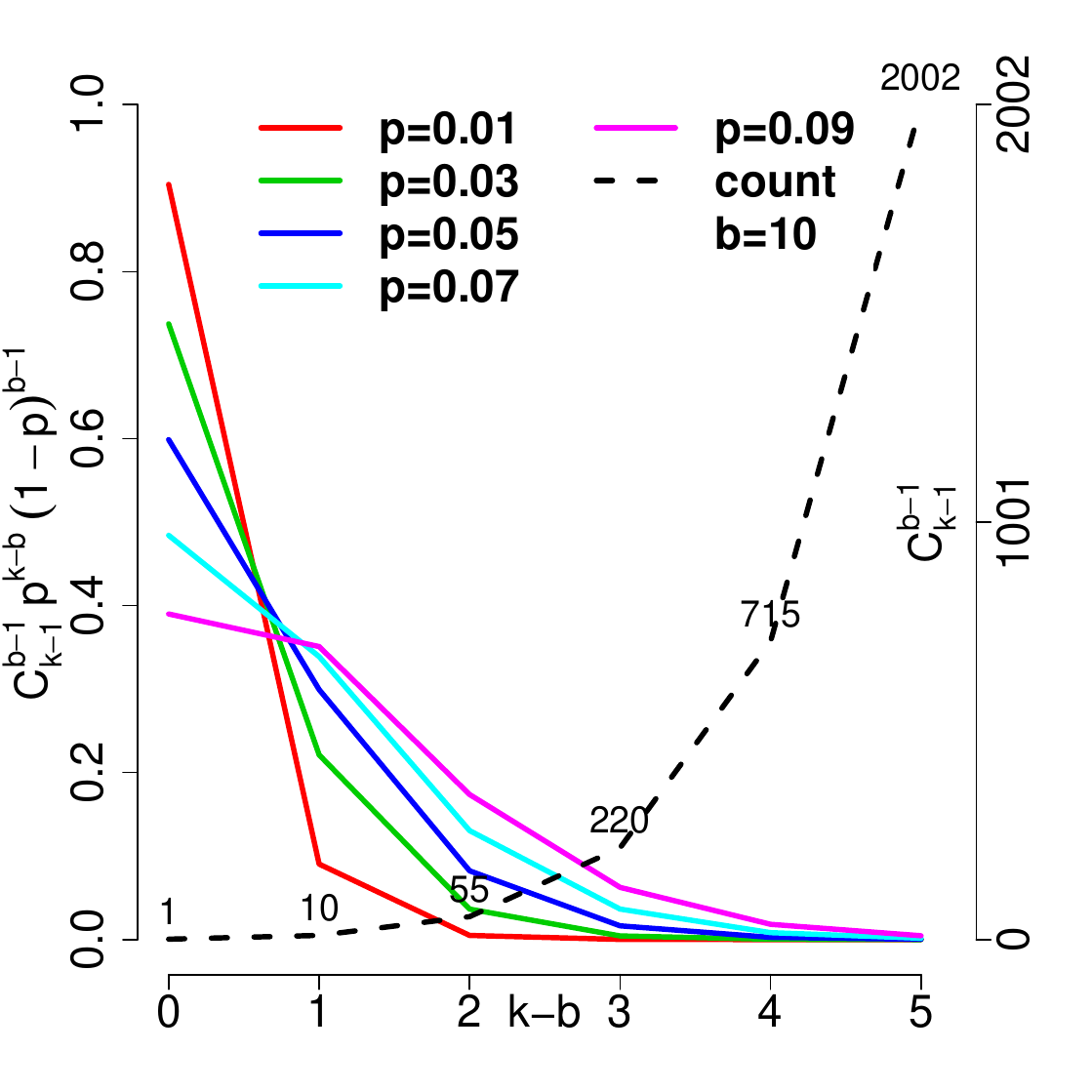}
\includegraphics[width=0.325\linewidth, height=0.32\linewidth, trim=0pt 0pt 1cm 0pt, clip]{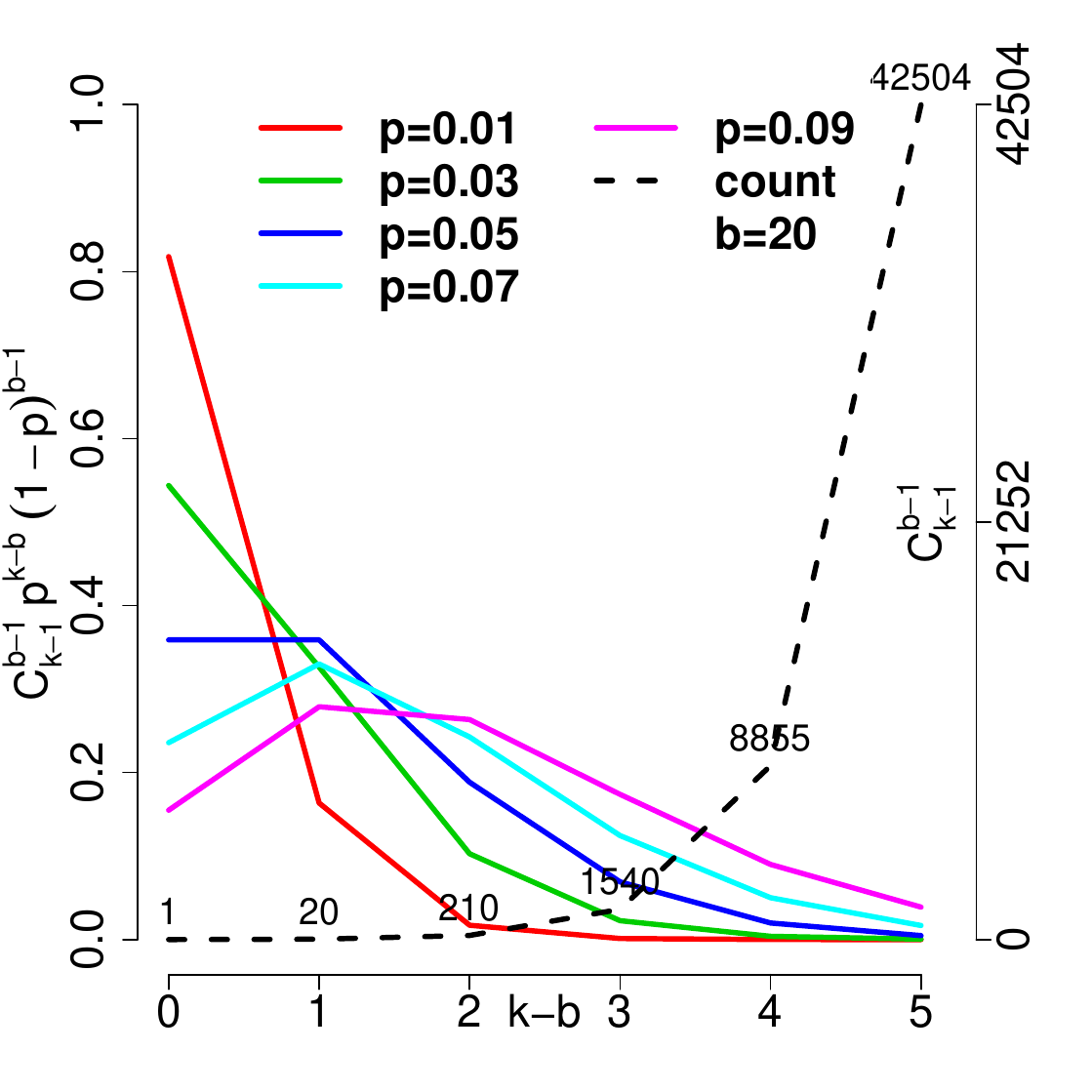}
\end{minipage}
\caption{Size of ensemble $\tilde{\mathcal{Y}}$ (right $y$-axis) and weights $w(k;p,b)$ assigned to ensembles of different sizes (left $y$-axis) for different $p$ and $b$}\label{fig:PRCO}. \vspace{-18pt}
\end{figure}

Figure \ref{fig:PRCO} provides a visual illustration on the ensemble effect achieved through the Bernoulli NI. The dashed line in each plot is associated with the right $y$-axis, which is the size (how many numbers) of an ensemble. The solid lines are associated with the left $y$-axis, which represent the weights assigned to ensembles of different sizes.
When there is no Bernoulli NI, $p=0$ and $k=b$, the ensemble is of size 1 (the first point on the dashed line in each plot). For $p>0$, we have more than one way of generating the set of $b$ observations; and the actual ensemble set size depends on $p$ and $k$.  In brief, for a fixed $b$, as $k$ increases, the size of the ensemble set $\tilde{\mathcal{Y}}$, $C_{k-1}^{b-1}$, increases dramatically (the dashed line within each plot), implying more sub-TS' are involved to obtain the posterior distribution of $s_i$. In addition, the ensemble set also increases dramatically with $b$ for a fixed $k-b$ value (the trend of the dashed lines across the 3 plots). The separated lines for different $p$ suggest that ensembles of different sizes are not weighted equally toward the conditional posterior distribution of $s_i$: the larger an ensemble, a smaller the  weight it carries, especially for small $p$. 
As stated in Sec \ref{sec:inference}, the goal of employing the Bernoulli NI is to create an ensemble of sub-TS' of considerable diversity among the ensemble members so to achieve more robust estimation.  Figure \ref{fig:PRCO} implies that  $p$ as small as $O(0.01)$ can create an ensemble of sub-TS' of enough diversity. A large $p$ leads to a larger ensemble, but it also implies higher computational cost which could overshadow the improved diversity. In addition, a large $p$ could also drop too many time points and lead to too much fluctuation in the sub-TS' from iteration to iteration, leading to possibly large bias or large variance in the estimation. Our empirical studies in Sec \ref{sec:simulation} explore further the effect $p$ on the estimation of state path. 


\subsubsection{Stability Improvement in Objective Function}\label{sec:stability}
The Bernoulli NI also improves the stability of the objective functions from which the MAP estimation is obtained in the presence of random perturbation in the TS. The formal result is presented in Proposition \ref{prop:perturb1}. The proof is given in the Appendix 
\begin{pro}[\textbf{improved stability of objective function}]\label{prop:perturb1}
Let $Y_i'=Y_i+z_i$, with $z_i\overset{\text{ind}}{\sim} N(0,\varepsilon^2)$ for $i=1,\ldots,n$, be an  externally perturbed observation to the original observation $Y_i$ from TS $\Y$; and $Y_i'$ comprises the  perturbed TS $\Y'$. Let $\tilde{\Y}$ and  $\tilde{\Y}'$ denote a sub-TS of $\Y$ and $\Y'$, respectively, after implementing the Bernoulli NI in an iteration of Algorithm \ref{alg:reSAVE}. The difference in the objection function (negative log-likelihood function or negative log-posterior distributions of $(\Theta,\bs\eta), S$ and $\bs\sigma^2$) given $\tilde{\Y}'$ vs. that  given $\tilde{\Y}$ after the Bernoulli NI is on average smaller than the difference obtained without the Bernoulli NI.
\end{pro}\vspace{-6pt}
Since the objective function is more stable with the Bernoulli NI than without in the presence of random external perturbations, the MLEs of $(\Theta,\bs\eta)$ and the MAP estimates of $(\Theta,\bs\eta,S,\bs\sigma^2)$ based on the former are expected to be more stable as well.  


\subsection{Foresting of Future State and Volatility}\label{sec:forecasting}
Foresting is often of major interest in TS analysis as they provide insights into future trends and are useful for decision making (e.g., developing option trading strategies in financial markets, predicting earthquakes). Algorithm \ref{alg:pred} lists the steps for the $h^{th}$-step-ahead prediction of future volatilities and states through a trained COMS-GARCH process.
\begin{algorithm}[!htb]
\caption{$h^{th}$-step-ahead prediction}\label{alg:pred}
\SetAlgoLined
\SetKwInOut{Input}{input}
\SetKwInOut{Output}{output}
\Input{$\hat{\bs\eta}_{\text{MAP}},\hat{\Theta}_{\text{MAP}},\hat{\sigma}_h^2$, MAP $S^*$ from Algorithm \ref{alg:reSAVE} given an observed TS of length $n$.}
\Output{Predicted future state path $S_{(n+1):(n+h)}$ and volatilities $\bar{\sigma}_{n+1}^2 ,\ldots,\bar{\sigma}_{n+h}^2$}
Define $S_n=\hat{S}^*$; and $\hat{\pi}_n\!=\!\hat{\Pr}(S_n=S^*)\!=\!1$\;
\For{$i= 1,\ldots, h$}{
Let $\hat{\bs{\sigma}}^2_{n+i},S_{n+i},\hat{\bs{\pi}}_{n+i}$ each be a $\nu^i\times 1$ vector\;
\For{$k= 1$ \textbf{to} $\nu^{i-1}$}{
Predict $\hat{Y}_{n+i}^2=\hat{\sigma}_{n+i-1}^2[k]\Delta t_{n+i}$; let $k'=S_{n+i-1}[k]$\;
\For{$j= 1$ \textbf{to} $\nu$}{
Calculate
\small $\!\hat{\sigma}_{n+i}^2[(k\!-\!1)\nu\!+\!j]\!=\!\hat{\alpha}(s_j) \Delta t_{n+i}\!+\!\!\left(\!\hat{\sigma}_{n+i-1}^2[k]\!+\!\hat{\lambda}(s_j)\hat{Y}_{n+i}^2\!\right)\exp(\!-\hat{\beta}(s_j)\Delta t_{n+i})$\;
Let $S_{n+i}[(k-1)\nu+j]=j$\;
Let $\small \hat{\Pr}(s_{n+i}\!=\!j\big|s_{n+i-1}\!=\!k',\hat{\bs\eta}_{
\text{MAP}})\!=\!\!
\begin{cases}
1\!-\!\exp(-\hat{\eta}_{\text{MAP},j,k'}\Delta t_{n+i})\mbox{ if }j\!\neq \!k'\\
2\!-\!\nu\!+\!\displaystyle\!\!\sum_{j'\ne k'}\!\exp(-\hat{\eta}_{\text{MAP},j',k'}\Delta t_{n+i})\mbox{ if }j\!=\!k'
\end{cases}\!\!\!\!\!\!$\;
Set $\hat{\bs{\pi}}_{n+i}[(k-1)\nu+j]=\hat{\bs{\pi}}_{n+i-1}[k]\hat{\Pr}(s_{n+i}=j|s_{n+i-1}=k',\hat{\bs\eta}_{\text{MAP}})$\;
}}
\small Predict volatility at $t_{n+i}:\bar{\sigma}_{n+i}^2\!=\! \hat{\bs{\pi}}^T_{n+i}\hat{\bs\sigma}^2_{n+i}$ and future path till $t_{n+i}:  S_{(n+1):(n+i)}\!=\!S_{n+i}\!\setminus\!S_n$.
}
\end{algorithm}\vspace{-18pt}

\section{Simulation Studies}\label{sec:simulation}
In this section, we demonstrate through simulation studies the applications of the COMS-GARCH process and illustrate the reSAVE algorithm in obtaining the MAP estimates of the model parameters, state path, and volatilities. 

\subsection{Simulation Study 2}\label{sec:sim1}
We use this simulation study to show the improvement in the robustness of  MAP estimates of volatilities  brought by the  Bernoulli NI in the reSAVE procedure in the COMS-GARCH process with one state.  
The COMS-GARCH process was simulated using the L\'{e}vy process as innovations \citep{COGARCH2004}, realized through the first-jump approximation that is normalized to have mean $0$ and variance $1$. We characterized the time gap $\Delta t$ between  two consecutive observations by a Poisson process with rate $\zeta$, that is, $\E(\Delta t)=\zeta^{-1}$; and examined $\zeta=2.5, 5, 10, 20$ (i.e., $\E(\Delta t)=0.4,0.2, 0.1, 0.05$, respectively).  The COMS-GARCH parameters are $\alpha=0.1\zeta,\beta=\log(10)\zeta,\lambda=\zeta$ 
The generated COMS-GARCH process is a discretized realization of the continuous-time process per Lemma \ref{lem:conv} as $\Delta t\rightarrow0$.  We generated 50 TS repetitions and set the number of time points in each simulated TS at $n=500$; we also examined several  Bernoulli NI rates  $p=0.03, 0.06, 0.09, 0.12$. To examine whether and how the Bernoulli NI help to improve the robustness of the volatility estimation, we perturbed each simulated TS by adding independently sampled noise from $N(0,0.1\!\cdot\!\mbox{SD}_{\Y})$ to each observation $Y_i$ for $i= 1,\ldots,n$ in TS $\Y$, where SD$_{\Y}$ is the sample standard deviation of $\Y$.

The COMS-GARCH model was fitted to the  the unperturbed and perturbed TS data in each repetition under each simulation scenario, following the steps of a simplified version of Algorithm \ref{alg:reSAVE} without the steps for state imputation or estimation of the transition parameters $\bs\eta$ as there is only one state. We imposed flat priors on $\Theta\!=\!(\alpha,\beta,\lambda)$. Per proposition \ref{prop:ensemble}, we used  $b=20$ observations $\tilde{Y}_i,\ldots, \tilde{Y}_{\min\{i+b,n\}}$ instead of the whole TS $\tilde{Y}$ when sampling $s_i$ from its full conditional posterior probability in each sub-TS to further save computational time.  The number of iterations $N$ was set at 300. 
The convergence of the reSAVE procedure was examined by visual inspection of the trace plots on the MAP estimates of $\Theta$. 

\begin{figure}[!htb] 
\begin{minipage}{1\textwidth}
\begin{center}
\includegraphics[trim=0.2cm 0 0.6cm 0, clip, height=0.33\textwidth]{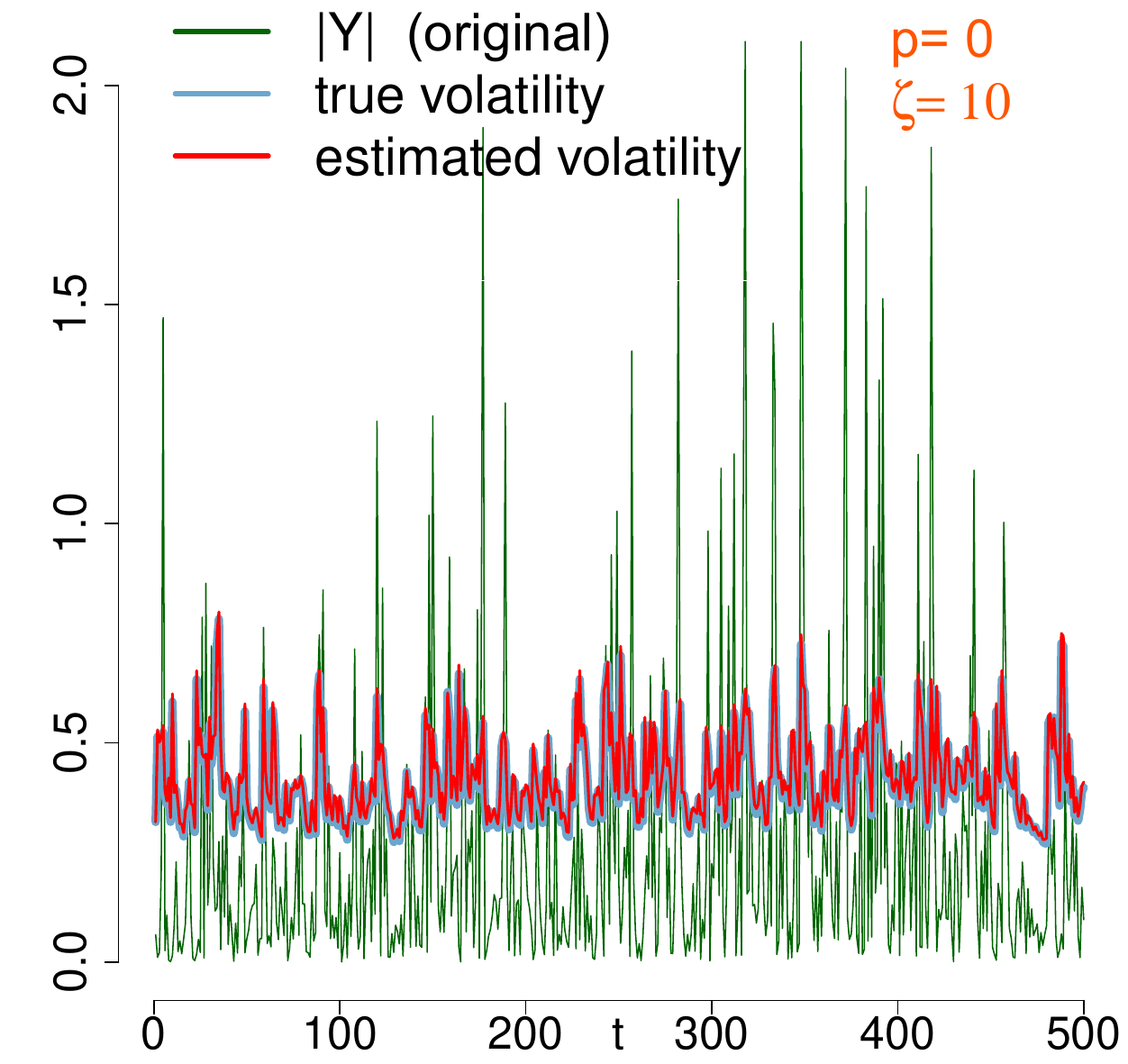}
\includegraphics[height=0.33\textwidth]{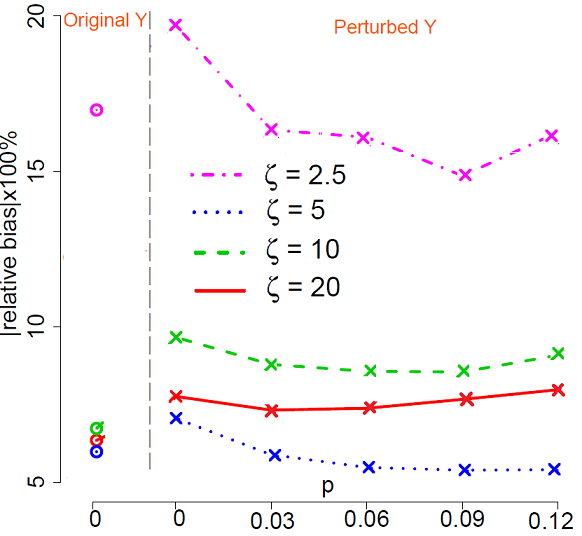}
\end{center}
\end{minipage}
\caption{Estimated volatility from one simulation repetition (left), and relative \%|bias| in the estimated volatility over 50 repetitions (right)}
\label{fig:sim1} \vspace{-12pt}
\end{figure}

Figure \ref{fig:sim1} presents an example on the estimated volatility superimposed on top of the true volatility from a randomly chosen repetition (left plot) and how much the Bernoulli NI can help reduce the relative \%|bias| of the estimated volatility ($l_1$-distance scaled by the true volatility and averaged over the 500 time points per TS and the 50 TS repetitions) (right plot).
In the left plot, the estimated volatilities almost completely overlap with the true volatility across all the time points, implying the reSAVE procedure performs well for this TS repetition. There are several observations from the right plot. First, the bias in the volatility estimation depends on  $\zeta$ and thus $\E(\Delta t)$. 
Second, the estimated volatilities can be sensitive to even mild fluctuation in the TS data as the bias with externally perturbed TS (crosses at $p=0$) increases compared to that at no external perturbation (circles at $p=0$). Third, the Bernoulli NI helps bring the bias down for all the examined $\zeta$ values, and in the case of $\zeta=2.5\; (\E(\Delta t)=0.4)$ and $\zeta=5\; (\E(\Delta t)=0.2)$, down to the level comparable to the unperturbed TS. There are not as much differences in the bias across $p$. 

\subsection{Simulation Study 2 }\label{sec:sim2}
In this simulation study, we demonstrate the inferential robustness and  computational efficiency of the proposed reSAVE procedure in a two-state COMS-GARCH process, compared the estimation vs that from the MSGARCH(1,1) process. 
We simulated 50 TS repetitions. In simulating each TS ($n=1,000$), we used the same the CO-GARCH process as the first simulation study, but separately in each of the two state.  Specifically, we characterized time gap $\Delta t$  by a Poisson process with rate $\zeta= 10, 40$, respectively. We set the GARCH model parameters at $\alpha(k)\!=\!\zeta c_k, \beta(k)\!=\!-\zeta\log(c_k), \lambda(k)\!=\!\zeta$ for  $k=1,2$. We let $c_2\!=\!2.5c_1$ and examined two values of $c_1$ at 0.1 and 0.025, respectively. The transitions of the states between two adjacent time points were modeled by a  hidden Markov process and the transition probabilities were calculated per Eqns (\ref{eqn:Sn1}) and (\ref{eqn:Sn2}) given the transition parameters $\bs{\eta}$ and simulated time gaps from the Poisson process. We examined two sets of $\bs{\eta}=(\eta_{12},\eta_{21})$ at (0.1, 0.1) and (0.25, 0.25), respectively. For the Bernoulli NI rate, we examined $p=0.01, 0.02, 0.03$. The externally perturbed TS data were obtained by adding independently sampled noise from $N(0,0.1\!\cdot\!\mbox{SD}_i)$ to each  $Y_i$ for $i\!=\! 1,\ldots,n$, where SD$_i$ is the sample SD of the subset of $\Y$ that are of the same state as $s_i$.

Algorithm \ref{alg:reSAVE} was applied to fit the COMS-GARCH model to each simulated TS in each simulation setting.  The number of iterations was set at 2,000 and the number of sampled paths $m$ was set at 6.  Per proposition \ref{prop:ensemble}, we used  $b=20$ observations $\tilde{Y}_i,\ldots, \tilde{Y}_{\min\{i+b,n\}}$ instead of the whole TS $\tilde{Y}$ when sampling $s_i$ from its full conditional posterior probability in each sub-TS to further save computational time. Non-informative flat priors were imposed on $\bs\theta$ and $\eta$. The MAP estimates of $(\Theta,\bs\eta)$ in each iteration were obtained directly via an optimizer.  The convergence of the algorithm was assessed by visual inspection of the trace plots on the MAP estimates of $(\Theta, \bs\eta)$ and the logarithm of the joint posterior distribution of $(\Theta,\bs\eta,\bs\sigma^2,S)$. The benchmark method MS-GARCH(1, 1) was fitted to the TS data using R package \texttt{MSGARCH}. Since the MS-GARCH model assumes evenly-space TS data, we first applied linear interpolation to each simulated TS to obtain the equally spaced TS data before the fitting the model.

\begin{figure}[!htb]
\hspace{1in} state path estimation \hspace{1.75in} volatility estimation\\
\includegraphics[trim=0.2cm 0 0 0, clip, width=0.245\linewidth]{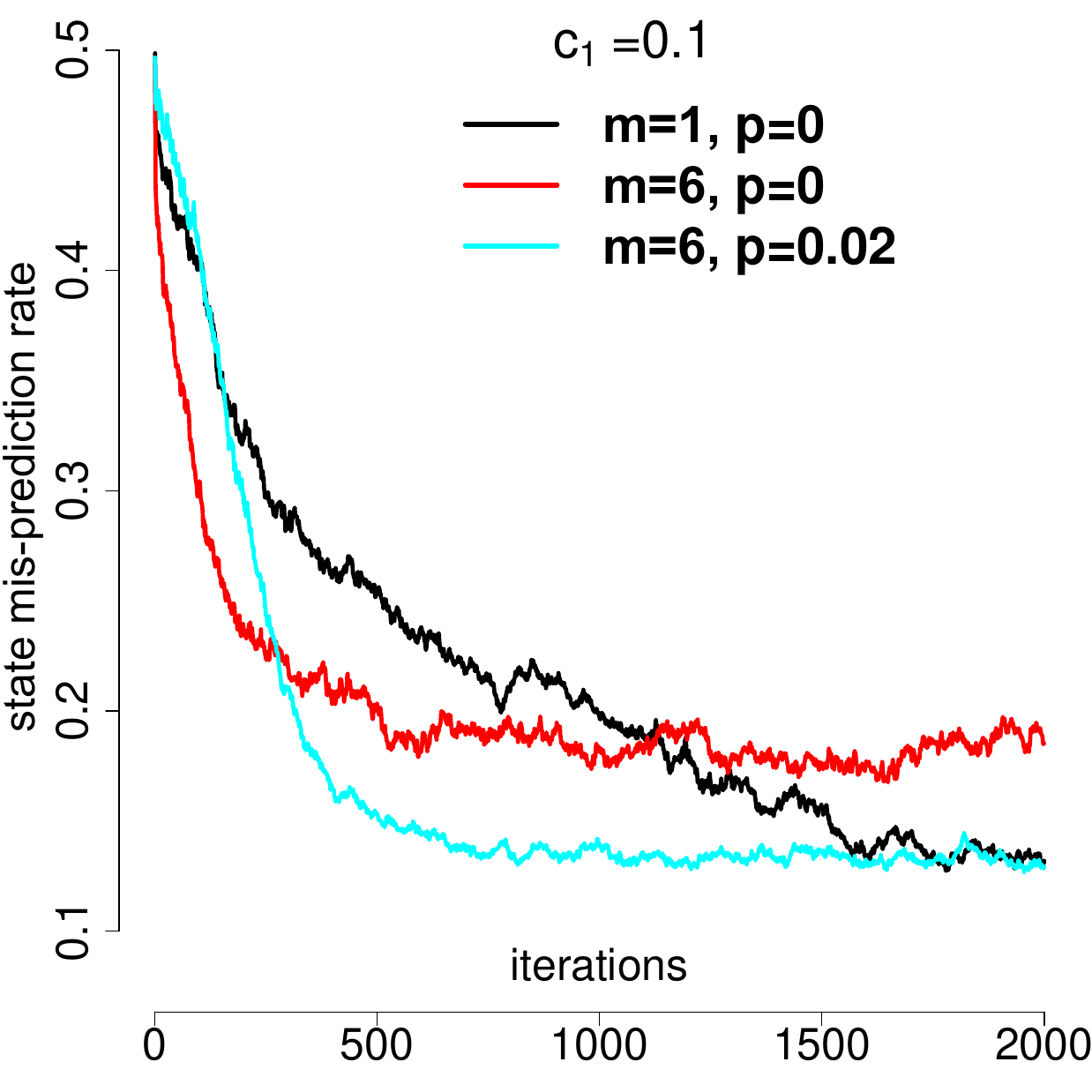}
\includegraphics[trim=0.2cm 0 0 0, clip, width=0.245\linewidth]{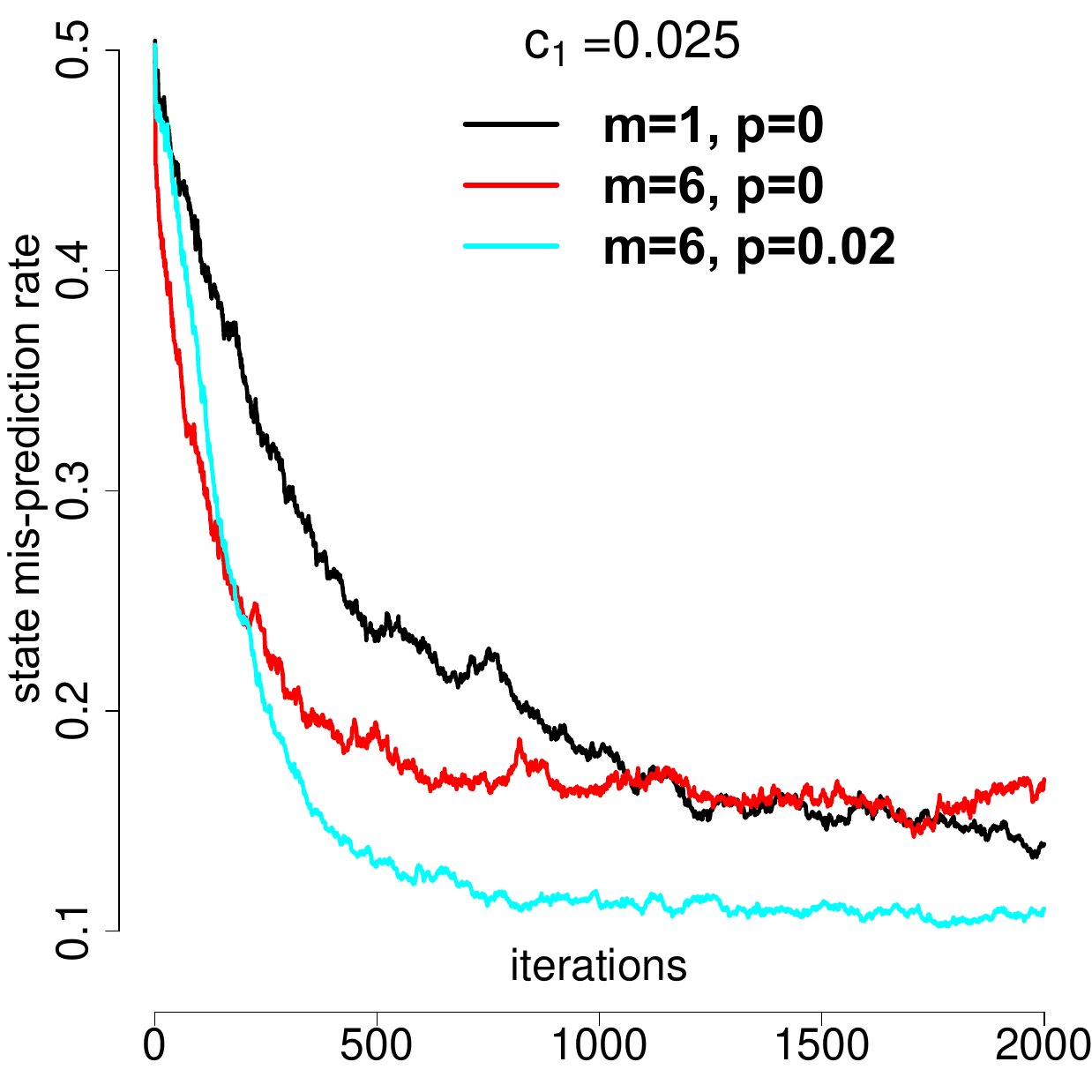}
\includegraphics[trim=0.2cm 0 0 0, clip, width=0.245\linewidth]{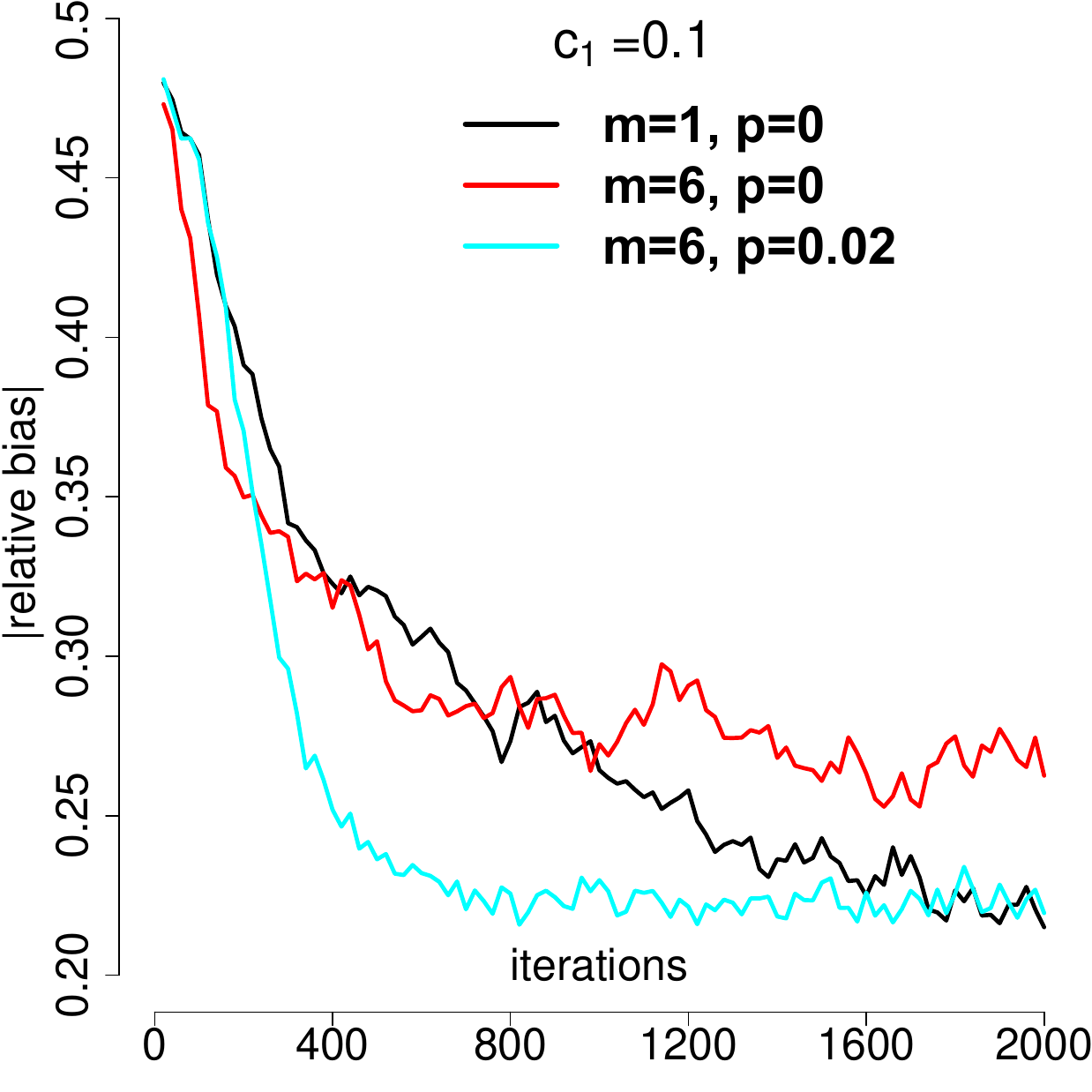}
\includegraphics[trim=0.2cm 0 0 0, clip, width=0.245\linewidth]{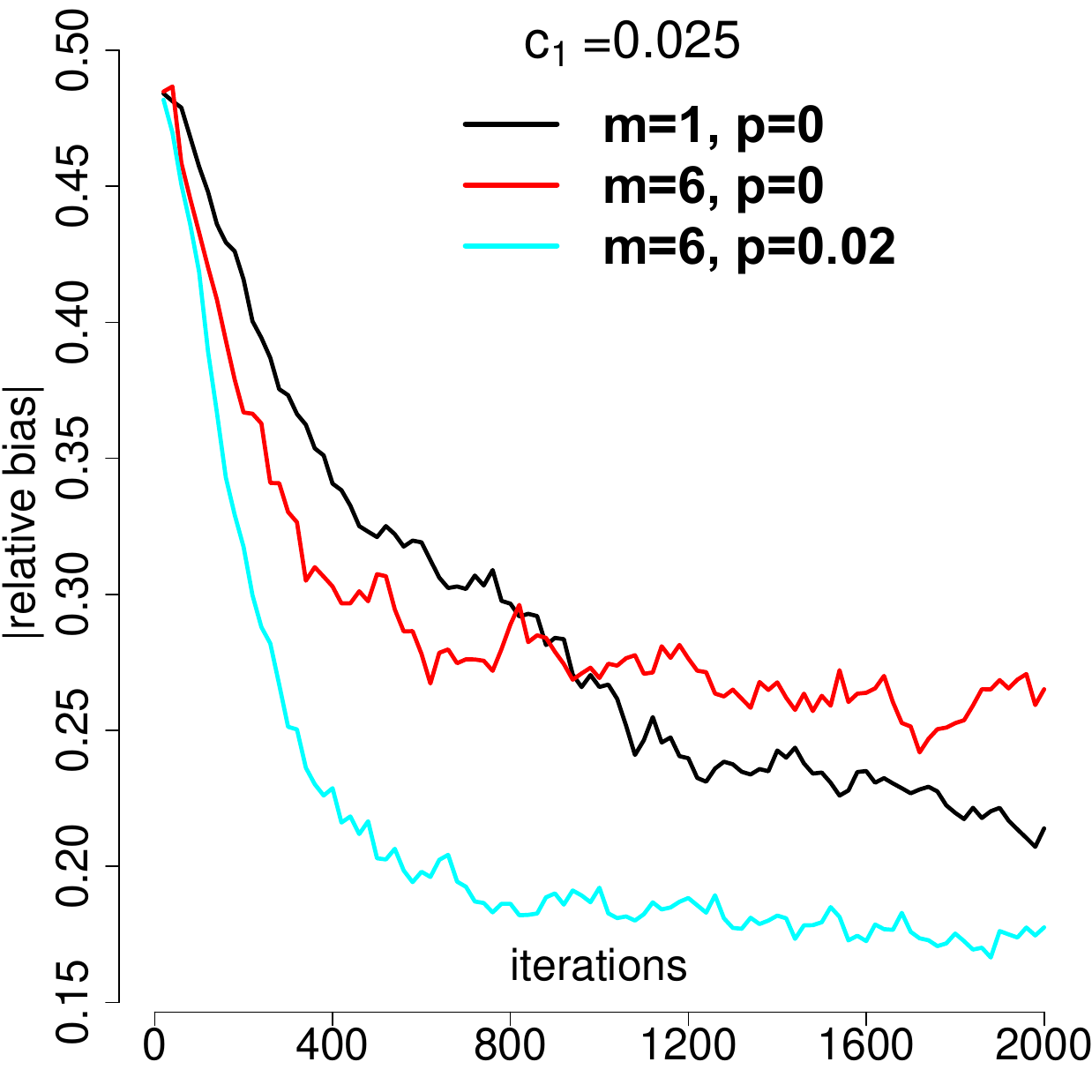}
\caption{Trace plots of state mis-prediction rate and |relative bias| of  estimated volatility in a single TS simulated at $\zeta=10,c_1=0.1,c_2=0.25, \eta_{12}=\eta_{21}=0.1$ } \label{fig:SIM3convergene}\vspace{-12pt}
\end{figure}

First, to illustrate the effects of the Bernoulli NI (with $b=20$) and the multiple path sampling scheme ($m=6$) in accelerating the convergence of the reSAVE procedure for estimating the MAP of the state path and volatilities, we compared the following three settings on convergence: 1) $m=6$ with Bernoulli NI at $p=0.02$ ($m=6,p=0.02$); 2) $m=6$ without Bernoulli NI ($m=6,p=0$); 3) $m=1$ (single path sampling) without Bernoulli NI ($m=1,p=0$). The results from 4 simulation scenarios are presented in Figure \ref{fig:SIM3convergene}. 
($m=6,p=0.02$) reached convergence within the least iterations ($\sim800$); ($m=6,p=0$) reached convergence around iteration $1,000$ in the state path estimation and around iteration $1,600$ in the volatility estimation. For ($m=1,p=0$), convergence is reached around $1,800$ for $c_1=0.1$, but does not seem to converge at iteration $2,000$.  In terms of the actual computational time for the reSAVE procedure, it is $O(Nd)$, where $N$ is the number of iteration and $d$ is the computational time per iteration. Though $b$ might be larger for $m>1$ with multi-path sampling and identifying the MAP path out of $m$, 
the longer per-iteration computational time in general does not overshadow the amount of time saved with the reduction in $N$, especially that $m$ is often a small number and parallel computing can used for sampling. In this simulation, the total computational time was less for $m=6$ on average  than for $m=1$. We provide in the Supplementary Materials some  examples on the MAP estimates for the state path and volatilities from a randomly selected TS analyzed in these three settings. It is obvious from the plot the setting of ($m=6,p=0.02$) has the most accurate estimation for the state path and volatilities across the time points.

\begin{figure}[!htb]
\footnotesize  \centering \textbf{(a) Original TS} \\
\raggedright COMS-GARCH w/ Bernoulli NI \hspace{0.75cm} COMS-GARCH w/o Bernoulli NI \hspace{1.75cm}  MS-GARCH\\ 
\includegraphics[trim=1.0cm 0 0.76cm 0, clip, width=0.32\textwidth]{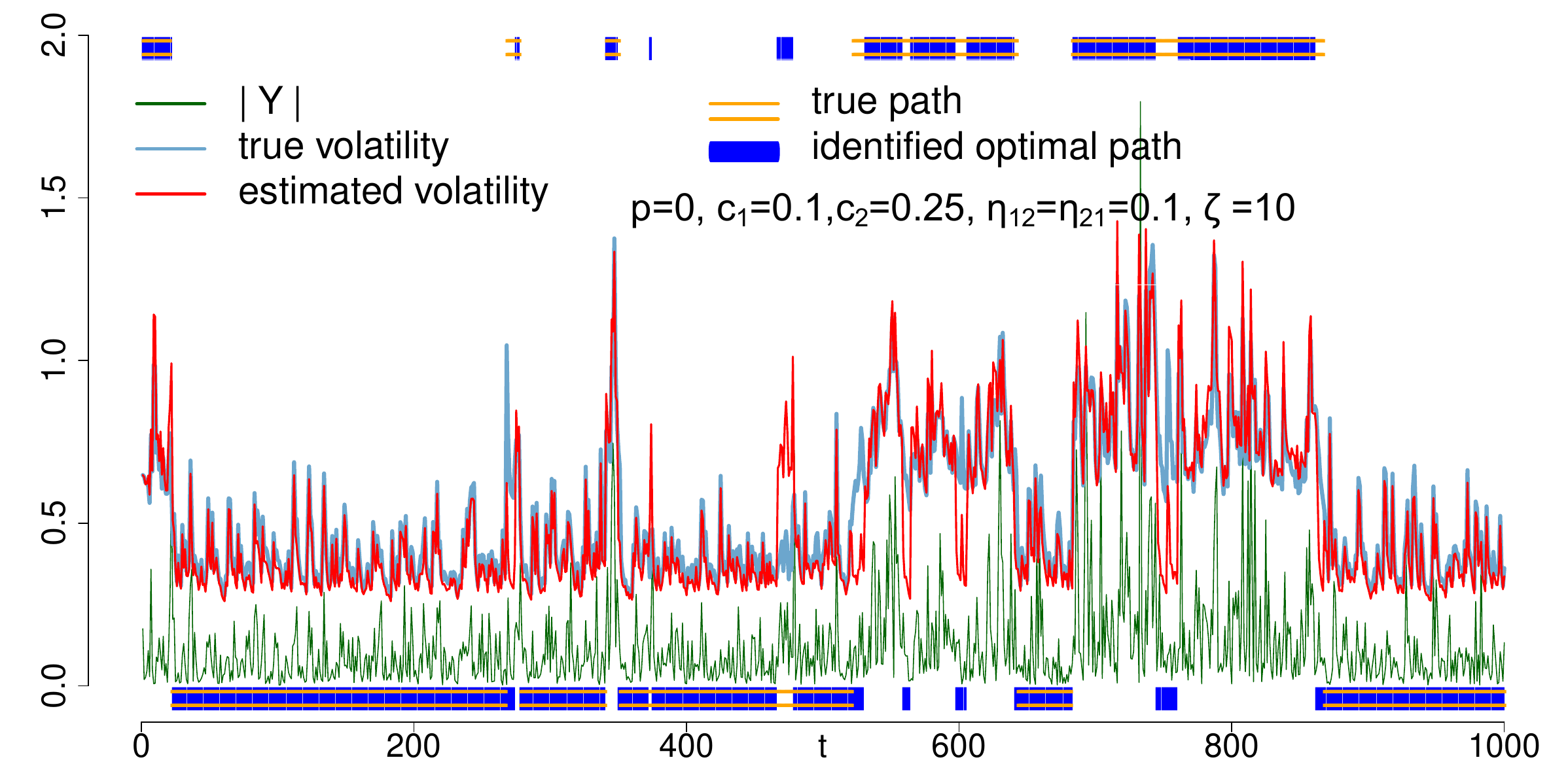}
\includegraphics[trim=1.0cm 0 0.76cm 0, clip, width=0.32\textwidth]{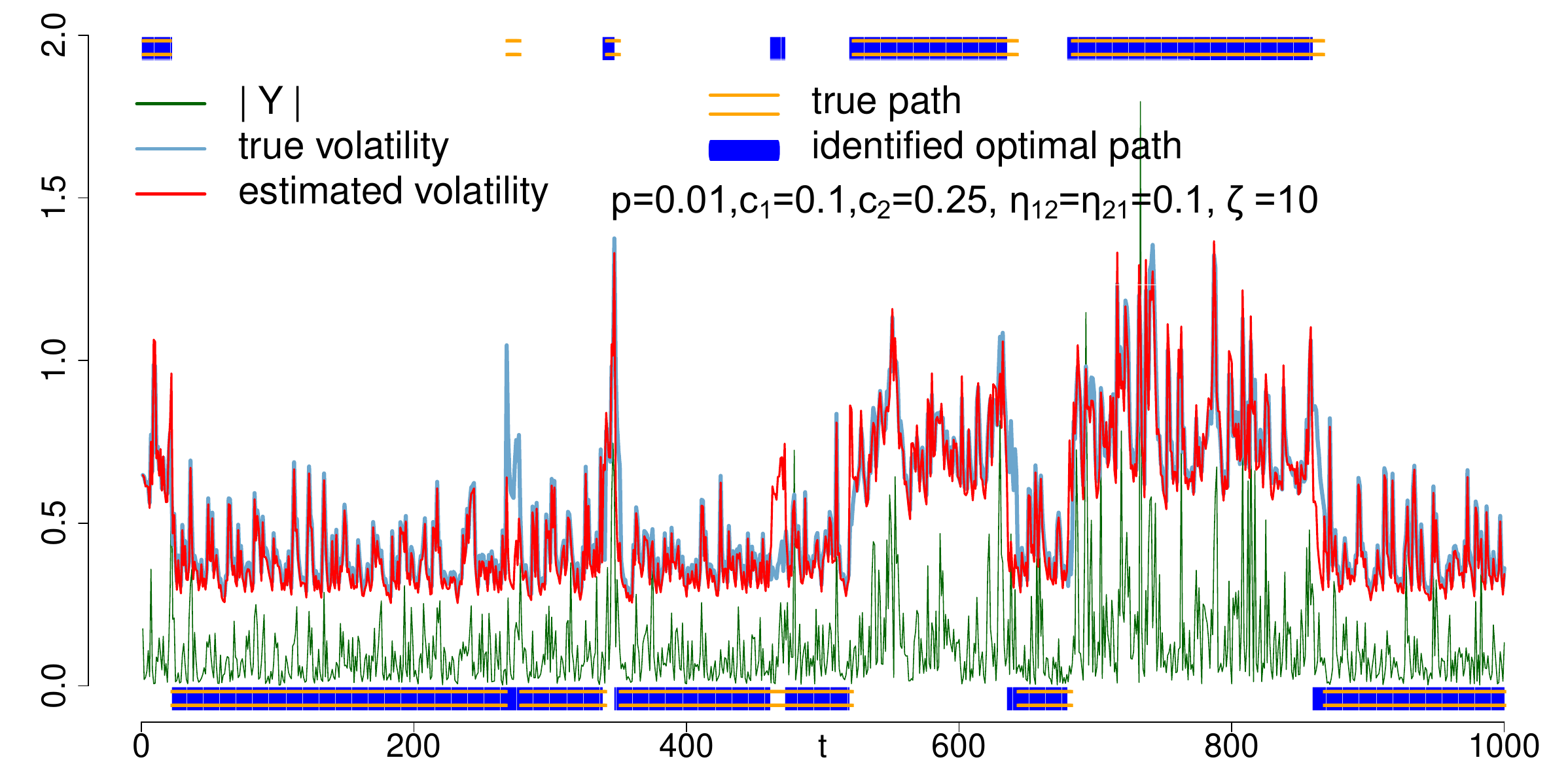}
\includegraphics[trim=1.0cm 0 0.76cm 0, clip, width=0.32\textwidth]{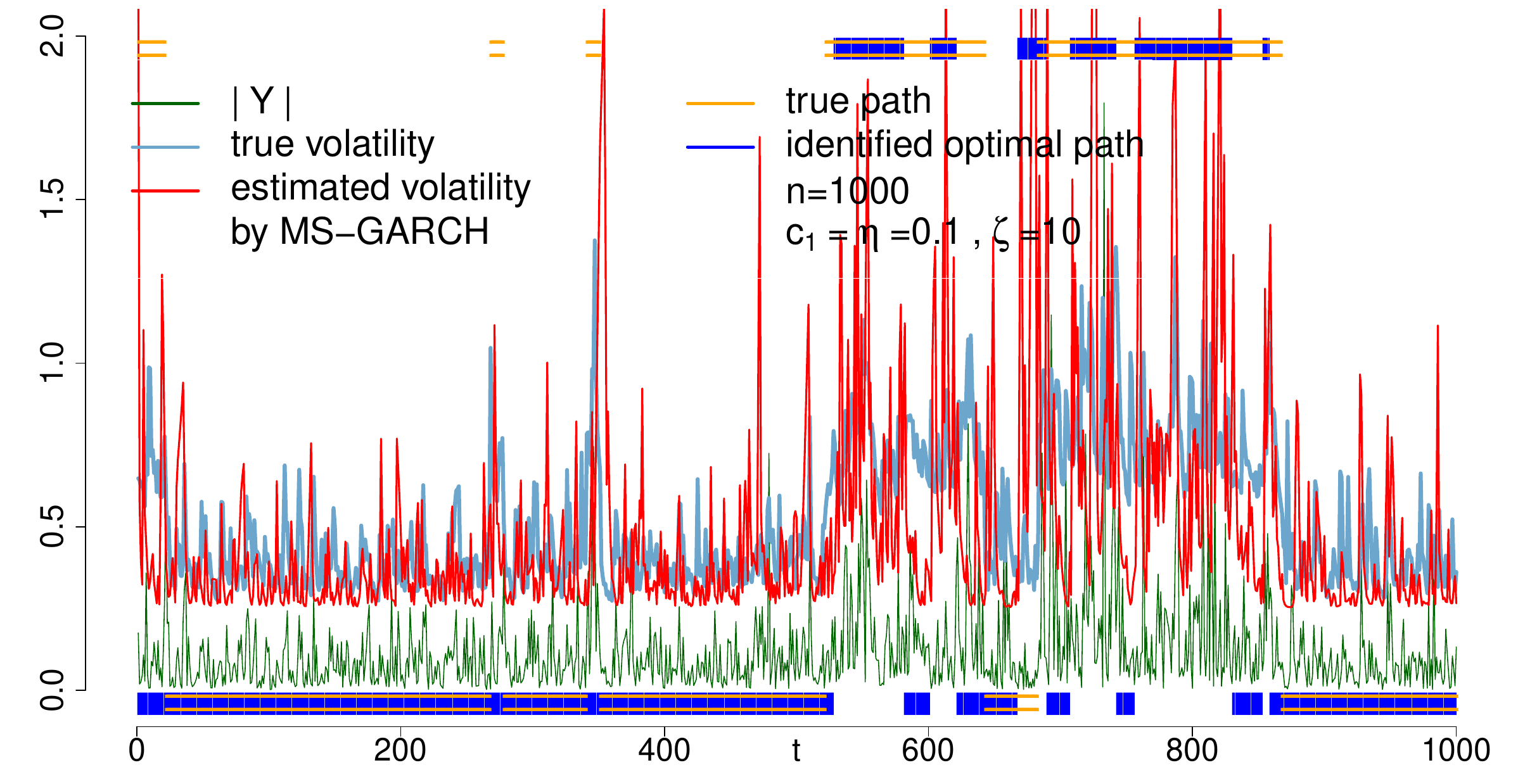}\\
\centering  \textbf{(b) perturbed TS} \\
\footnotesize COMS-GARCH w/ Bernoulli NI \hspace{1.5cm} COMS-GARCH w/o Bernoulli NI 
\includegraphics[trim=1.05cm 0 0.76cm 0, clip, width=0.49\linewidth]{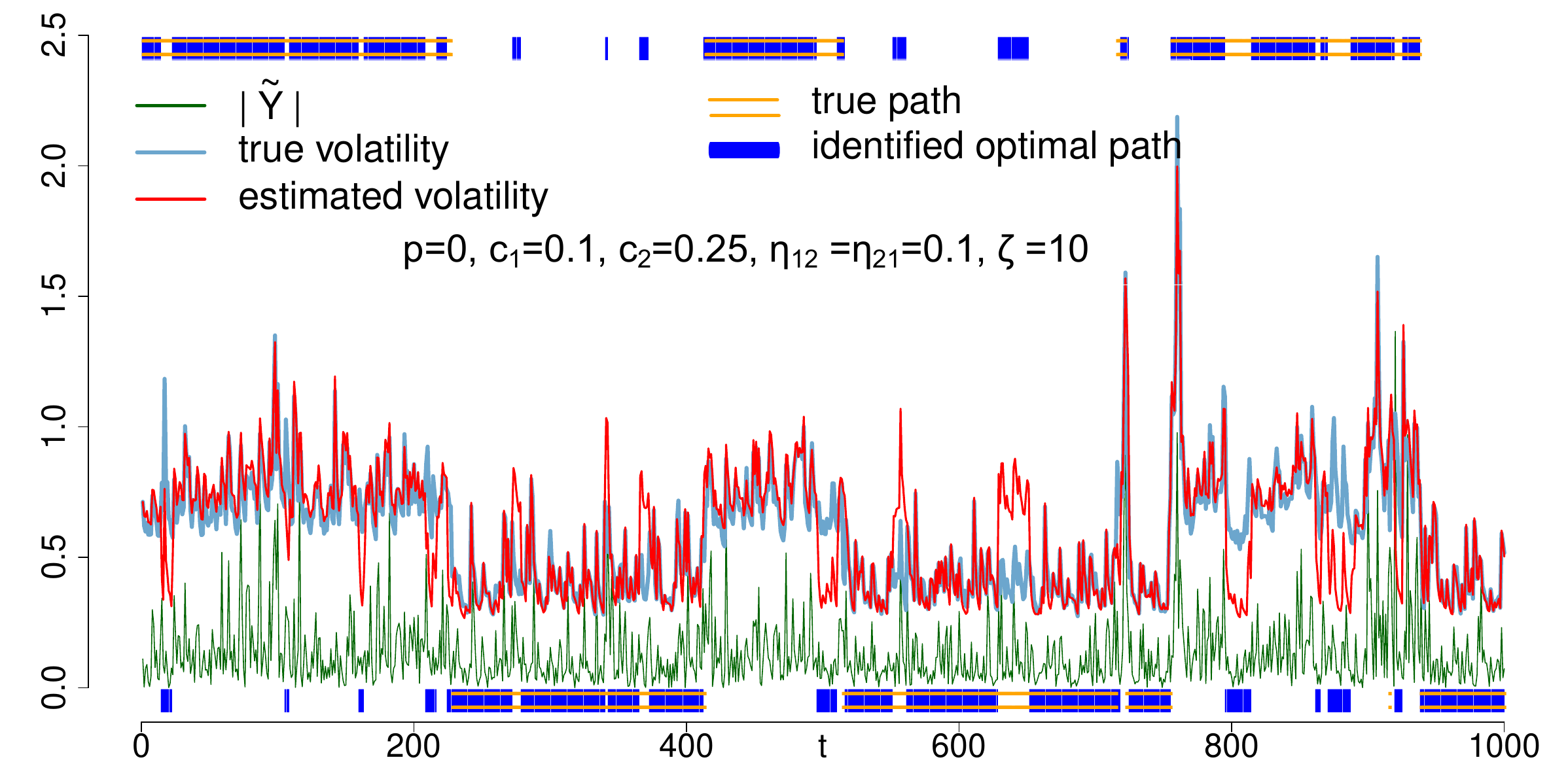}
\includegraphics[trim=1.1cm 0 0.76cm 0, clip, width=0.49\linewidth]{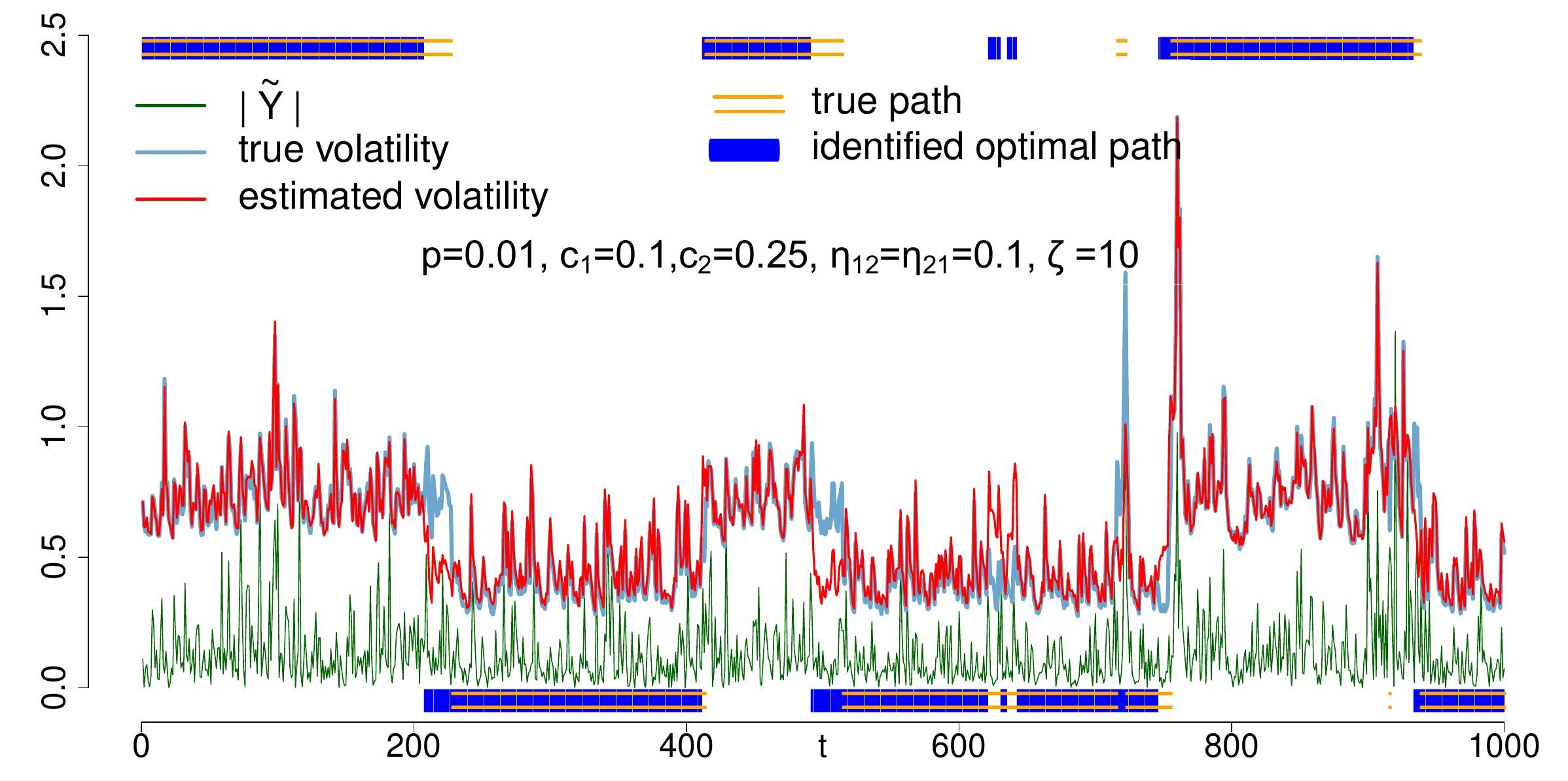}
\caption{Examples of predicted volatility and state path in a TS simulated at $c_1=0.1,c_2=0.25,\eta_{12}=\eta_{21}=0.1,\zeta=10$  ($m=6$ in the reSAVE procedure for COMS-GARCH)}\label{fig:SIM2eg}\vspace{-12pt}
\end{figure}

We provide in Figure \ref{fig:SIM2eg} an example on the MAP estimates via the reSAVE procedure  on the state path and volatilities from a randomly selected TS simulated at $\zeta=10, c_1=0.1, \eta_{12}=\eta_{21}=0.1$ with Bernoulli NI ($p=0.02$) vs.~without, along with the  benchmark results from the  MS-GARCH model. When there is no external perturbation in the TS, the Bernoulli NI does not negatively affect the volatility and state estimation; when there is, the Bernoulli NI improves the accuracy in the state and  volatility estimation. This observation suggests that the Bernoulli NI is a ``intelligent'' technique, only doing its tricks when needed and is silent otherwise. 
As for the MS-GARCH model, the volatilities tend to be over-estimated and there are more mis-predicted states compared to using the COMS-GARCH process.

\begin{figure}[!htb]
\textcolor{orange}{\textbf{\small 
\hspace{2cm}state mis-prediction rate \hspace{2cm} relative \%|bias| of estimated volatility}}\\
\centering\textcolor{blue}{\textbf{\small Original TS}}\\
\includegraphics[width=0.24\linewidth]{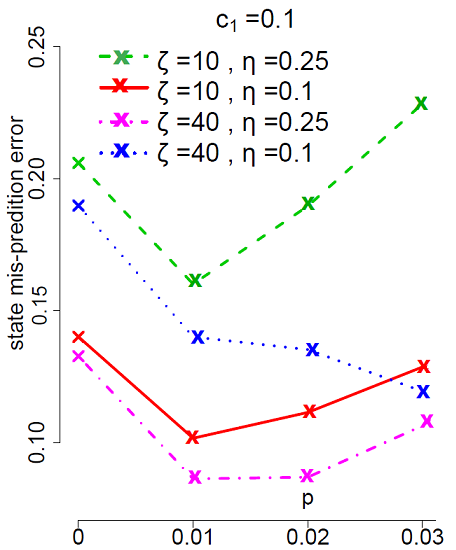}
\includegraphics[width=0.24\linewidth]{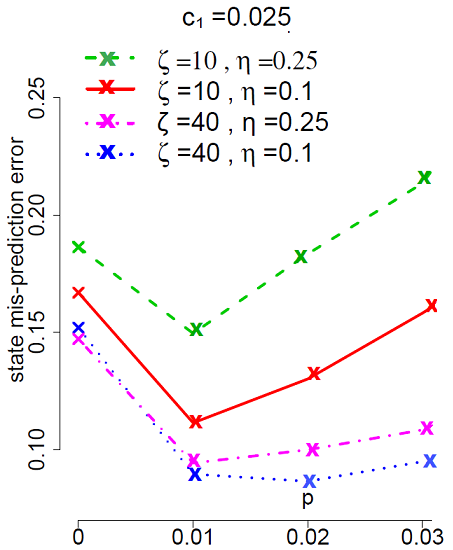}
\includegraphics[width=0.24\linewidth]{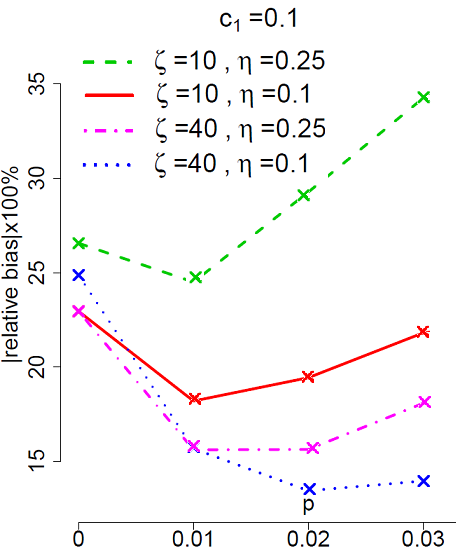}
\includegraphics[width=0.24\linewidth]{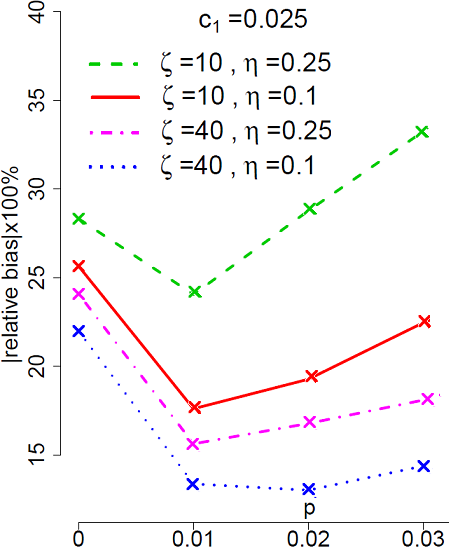}
\centering\textcolor{blue}{\textbf{\small Perturbed TS}}\\
\includegraphics[width=0.24\linewidth]{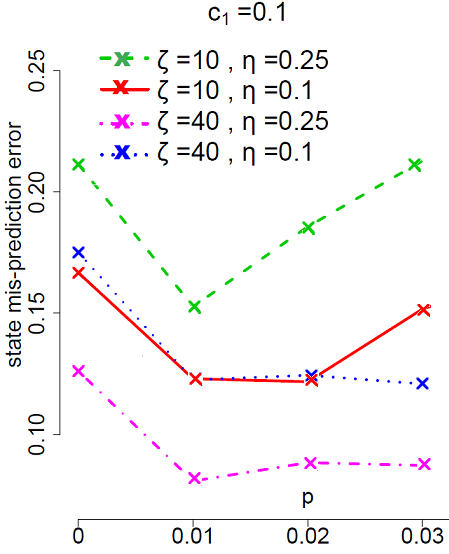}
\includegraphics[width=0.24\linewidth]{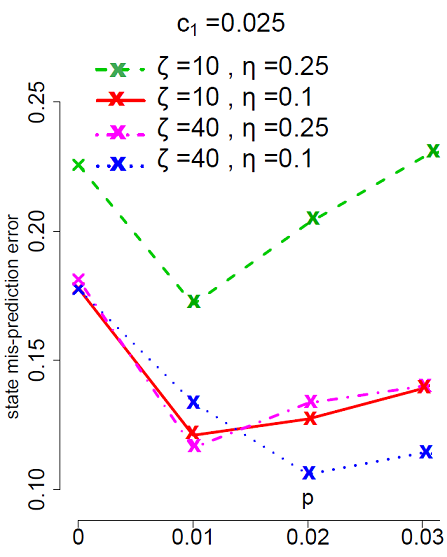}
\includegraphics[width=0.24\linewidth]{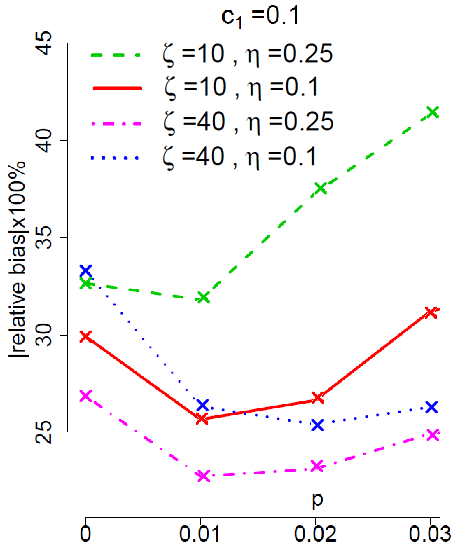}
\includegraphics[width=0.24\linewidth]{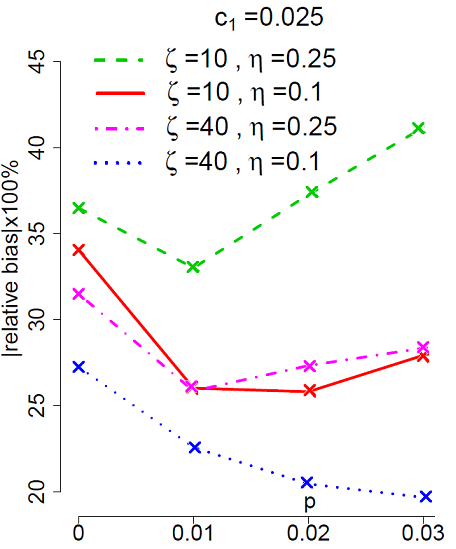}
\caption{State mis-prediction rate (state estimation bias) and volatility estimation bias vs Bernoulli NI rate $p$ in COMS-GARCH model ($m=6$ in the reSAVE procedure)}
\label{fig:SIM2no}\vspace{-15pt}
\end{figure}

To examine the effectiveness of the reSAVE procedure in improving the state path and volatility estimation, we summarized in Figure \ref{fig:SIM2no} the state mis-prediction rate and the relative \%|bias| of the estimated volatilities over the 1,000 time points per TS and the 50 TS repetitions . 
In all the examined simulation scenarios, the accuracy of the path identification and volatility estimation is significantly improved with a proper Bernoulli NI rate $p$ than without NI. 
The smallest mis-prediction rate (8\% $\sim$ 17\%) and volatility estimation bias (12\% $\sim$ 25\%)  are achieved around $p$ at $0.01\sim 0.02$ in most scenarios; and further increasing $p$ does not seem to improve the prediction accuracy, implying that selecting a proper choice of $p$ -- such as via the CV procedure in Algorithm \ref{alg:cv} -- is important. In addition, how much the Bernoulli NI helps in reducing the prediction bias relates to $\zeta$ (gaps between two adjacent time points), $\bs\eta$ (transition parameters), and the values of $\Theta$. 

\begin{figure}[!htb]
\centering
\includegraphics[width=0.85\linewidth,height=0.4\linewidth]{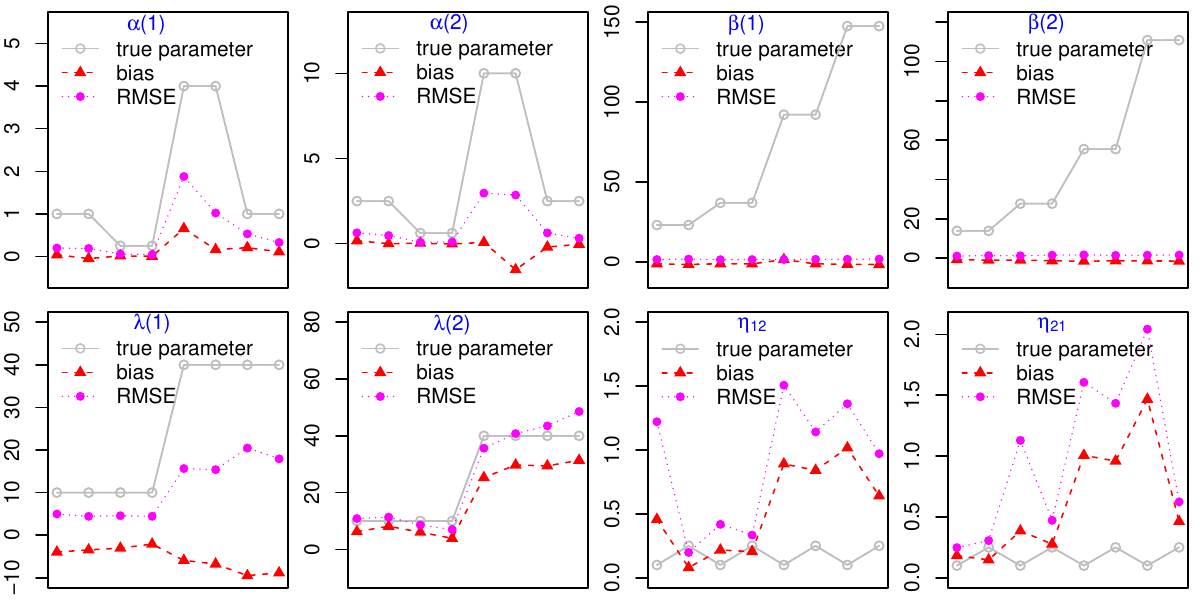}
\caption{MAP estimates and RMSE for COMS-GARCH  Parameters $\Theta$ and $\bs\eta$ via the reSAVE procedure with $p=0.01$ and $m=6$. $x$-axis the represents the eight COMS-GARCH model scenarios (defined by two $\eta$ values, two $c_1$ values, and two $\eta_{12}=\eta_{21}$ values)} \label{fig:biasMSE}\vspace{-24pt}
\end{figure}

The biases and root mean squared errors (MSE) for the MAP estimates of COMS-GARCH parameters $\Theta$ and $\bs\eta$ are presented in Figure \ref{fig:biasMSE}.  The estimates for $\alpha(k)$ and $\beta(k)$ are generally accurate and are noticeable for some simulation scenarios for $\lambda(k)$ and $\bs\eta$. The relatively large bias for $\bs\eta$ can be at least partially attributed to the low transition probabilities between different states, leading to data sparsity in estimating $\bs\eta$ (we provide in the Supplementary Materials the histograms of $\Pr(s_1|s_2)=\Pr(s_2|s_1)$ for $\eta_{12}=\eta_{21}=0.1$ and 0.25, respectively. In both cases, the mode transition probabilities is close to 0). Regarding the estimation of $\Theta$, the existence of bias in the GARCH, CO-GARCH and MS-GARCH parameters is rather a common problem than something unique to the Gibbs sampler or the reSAVE procedure we propose for the COMS-GARCH process. \citet{bollerslev1994arch} suggested that the MLE for the parameters from the GARCH model are biased; the estimation bias for the GARCH parameters for both the quasi-MLE and the constrained M-estimators (more robust) can be as large as 20\% in the empirical studies in \citet{melo2000assessing}; that for the parameters for the CO-GARCH process in \citet{COGARCHMLE2008} can be as large as 30\%. How to improve the statistical accuracy for the model parameters is on our future research agenda (more details are provided in Sec \ref{sec:discussion}).   
\section{Applications}\label{sec:case}
\begin{figure}[!htb]
\centering
\includegraphics[width=0.32\linewidth, height=0.245\linewidth,trim=0 0 1cm 0, clip] {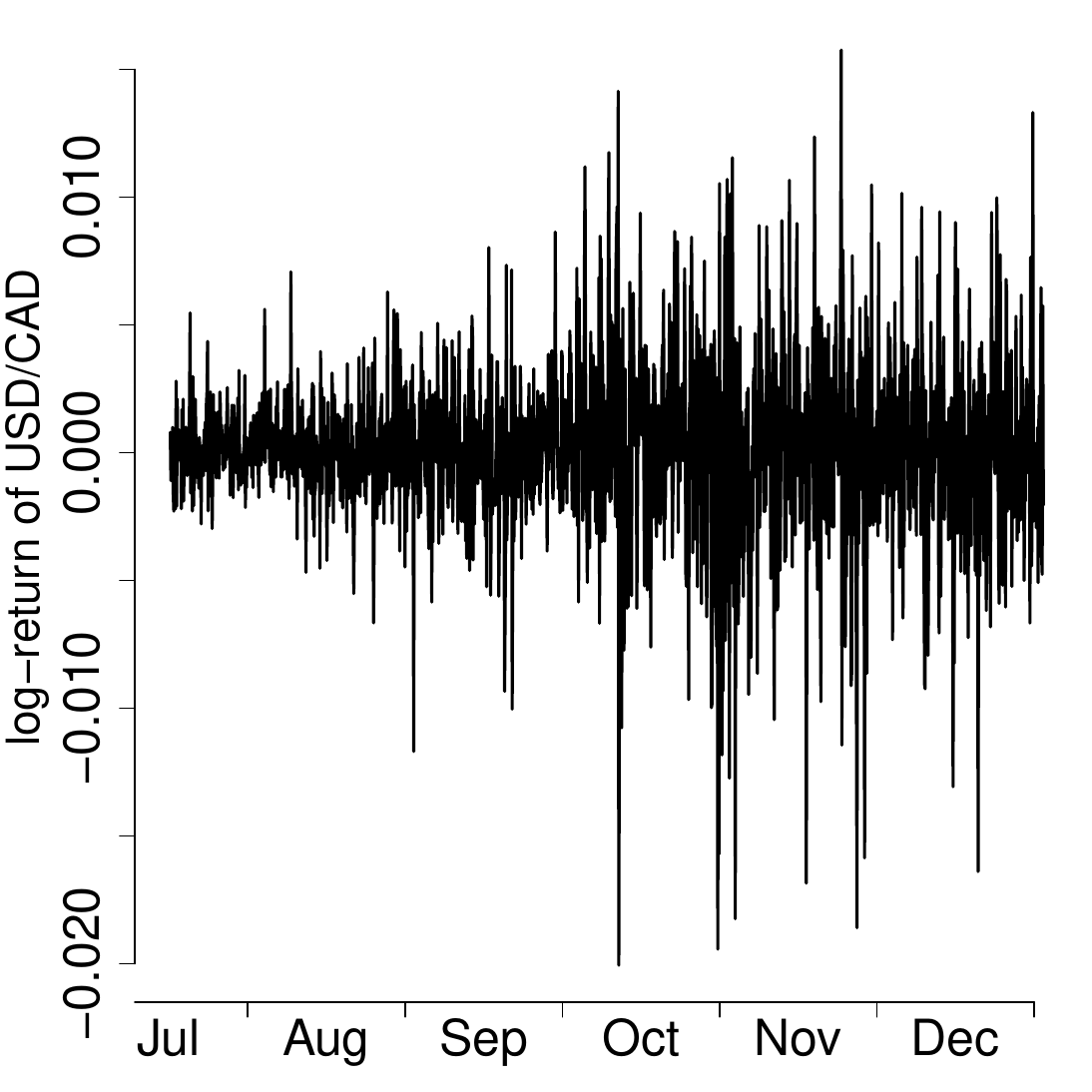}
\includegraphics[width=0.32\linewidth, height=0.245\linewidth,trim=0 0 1cm 0, clip] {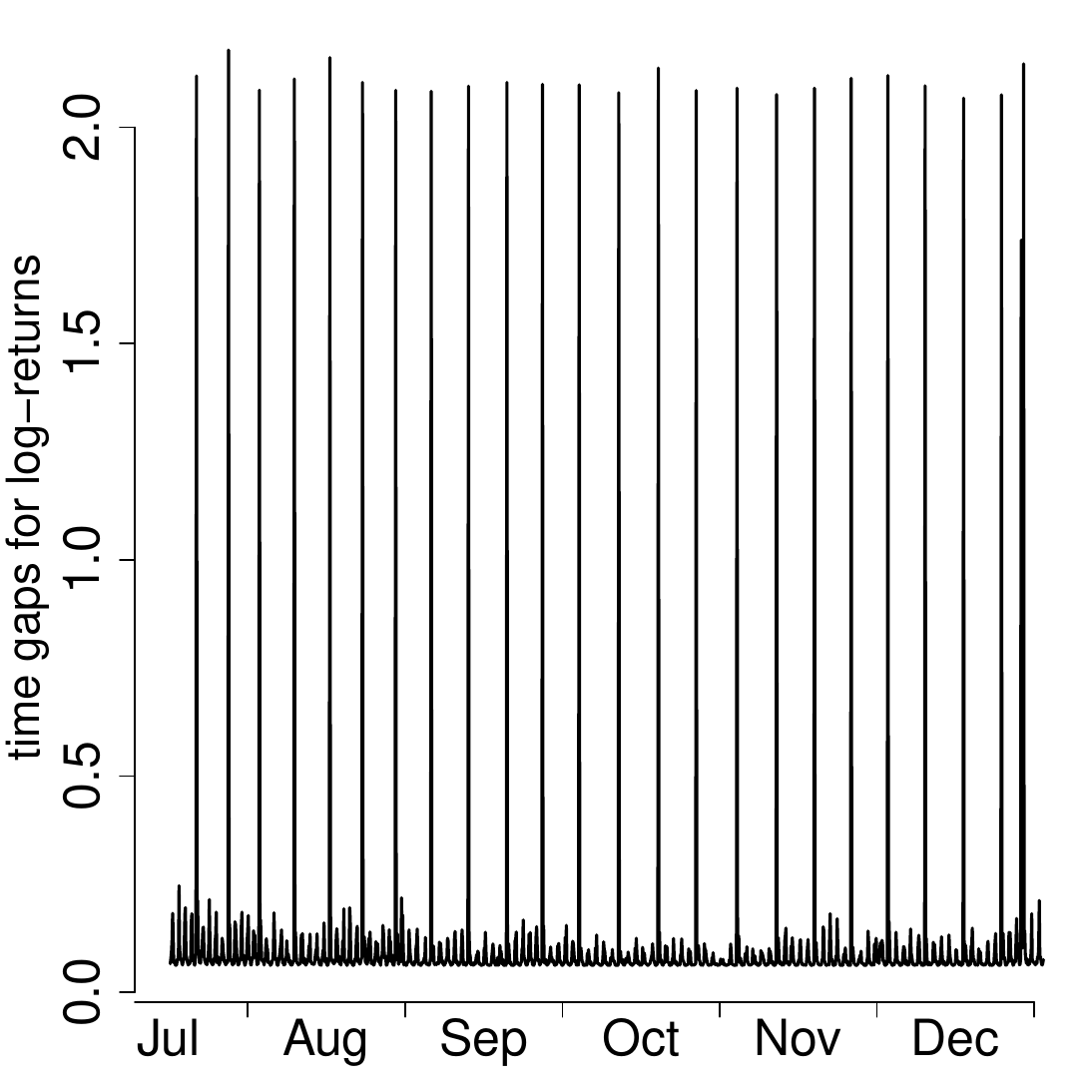} 
\includegraphics[width=0.32\linewidth, height=0.24\linewidth,trim=0 0 2cm 0, clip] {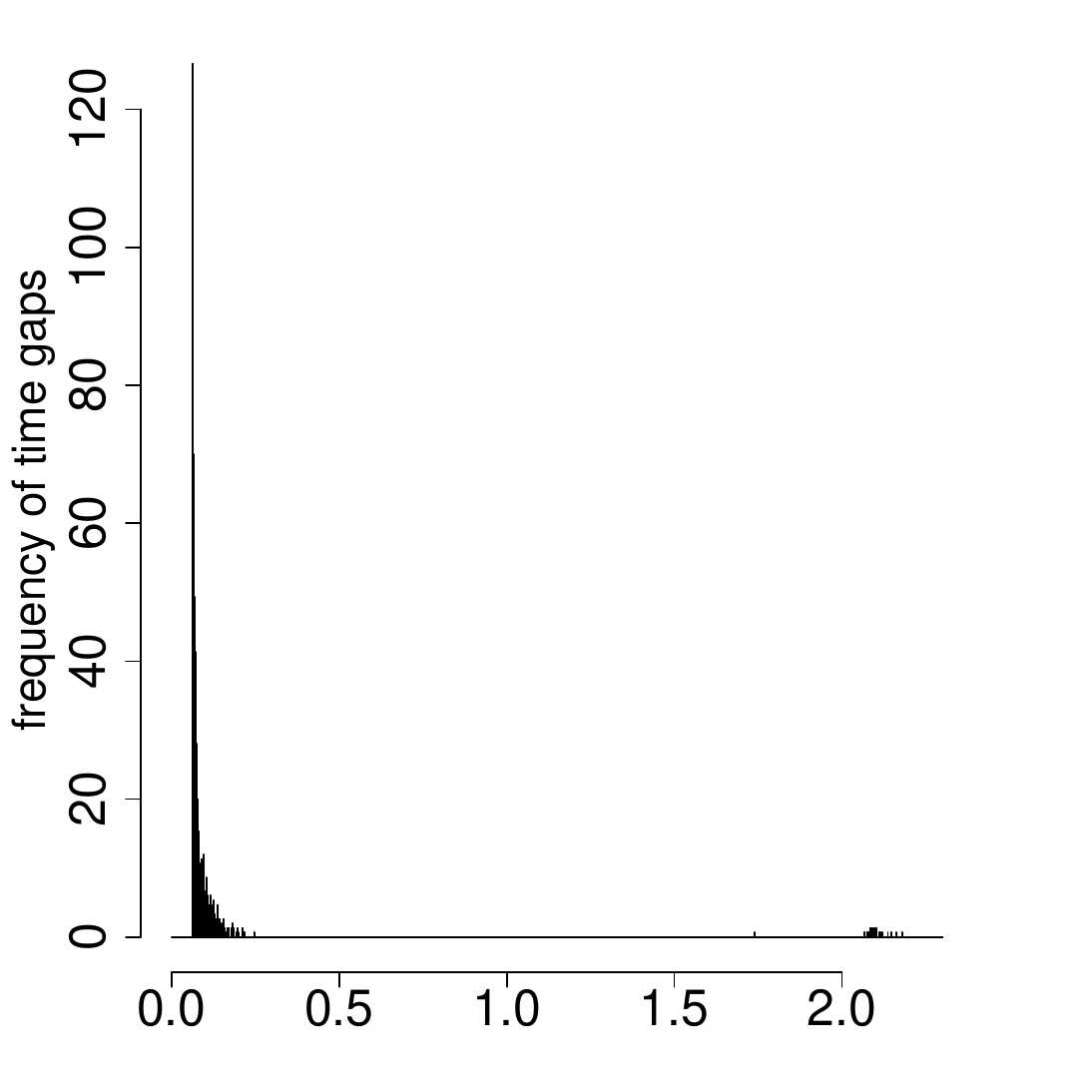}\\
\vspace{-3pt} (a) exchange rate TS\\
\includegraphics[width=0.32\linewidth, trim=0 0 0cm 0, clip] {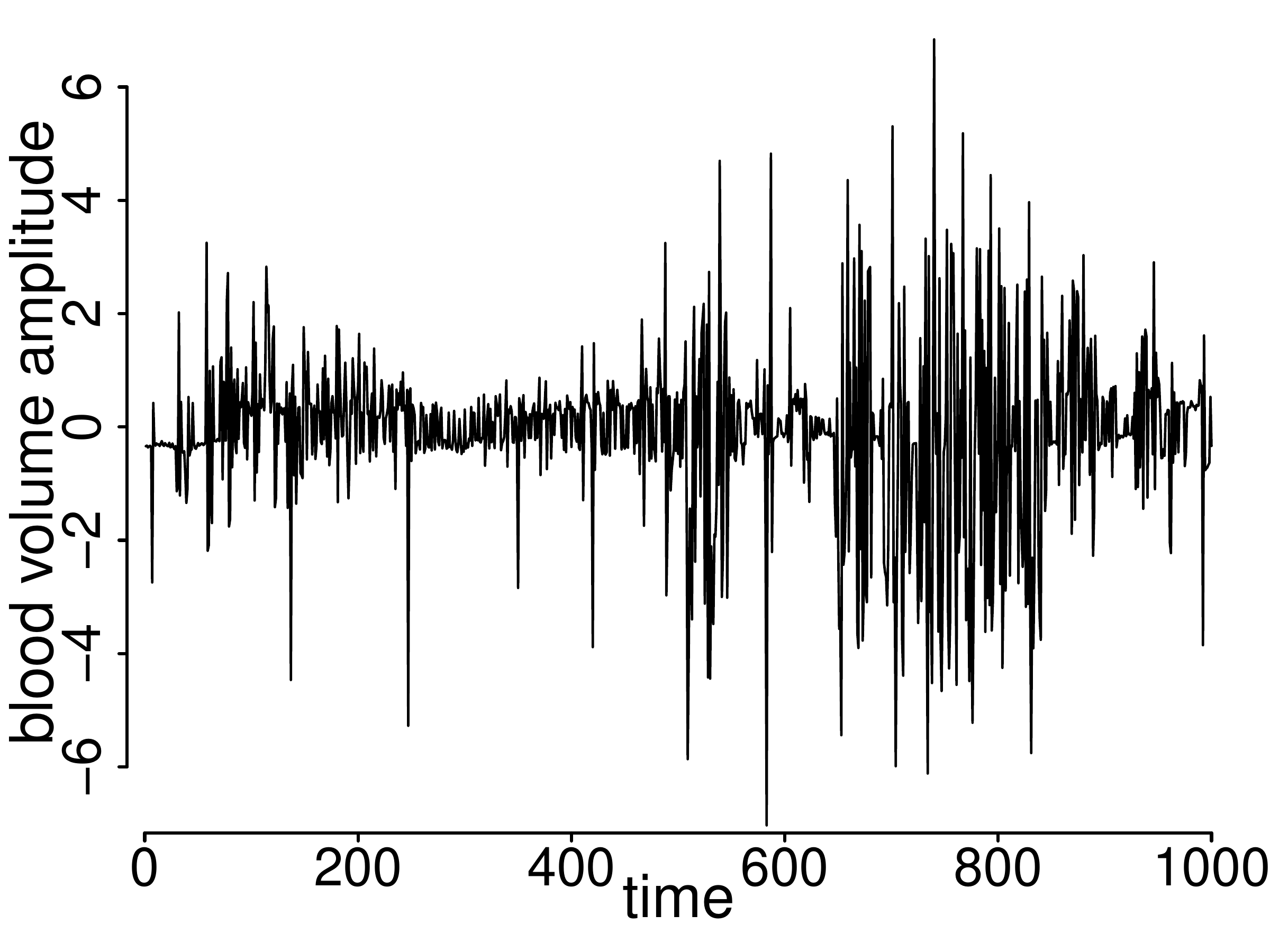}
\includegraphics[width=0.32\linewidth, trim=0 0 1cm 0, clip] {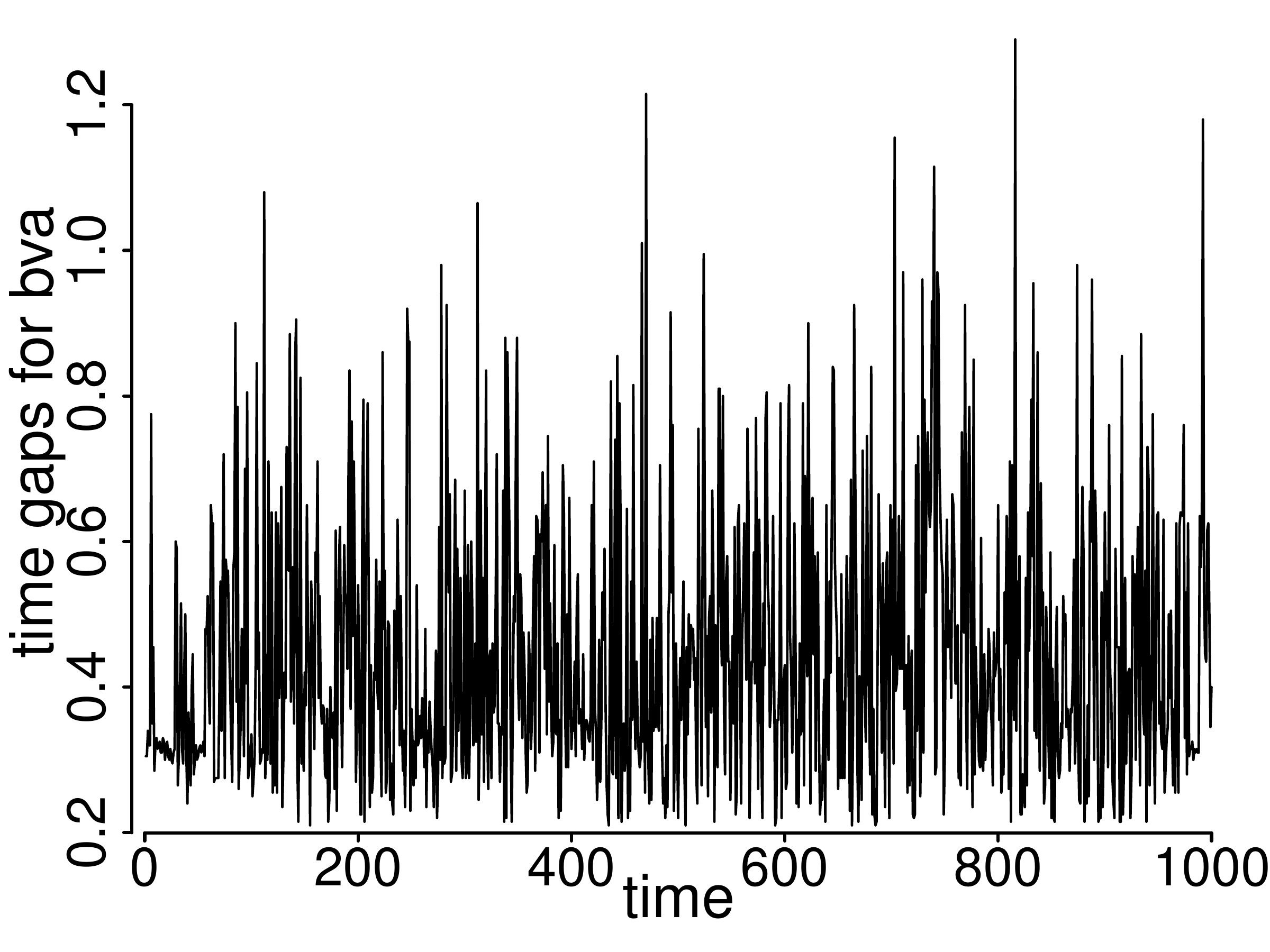}
\includegraphics[width=0.32\linewidth, trim=0 0 2cm 0, clip] {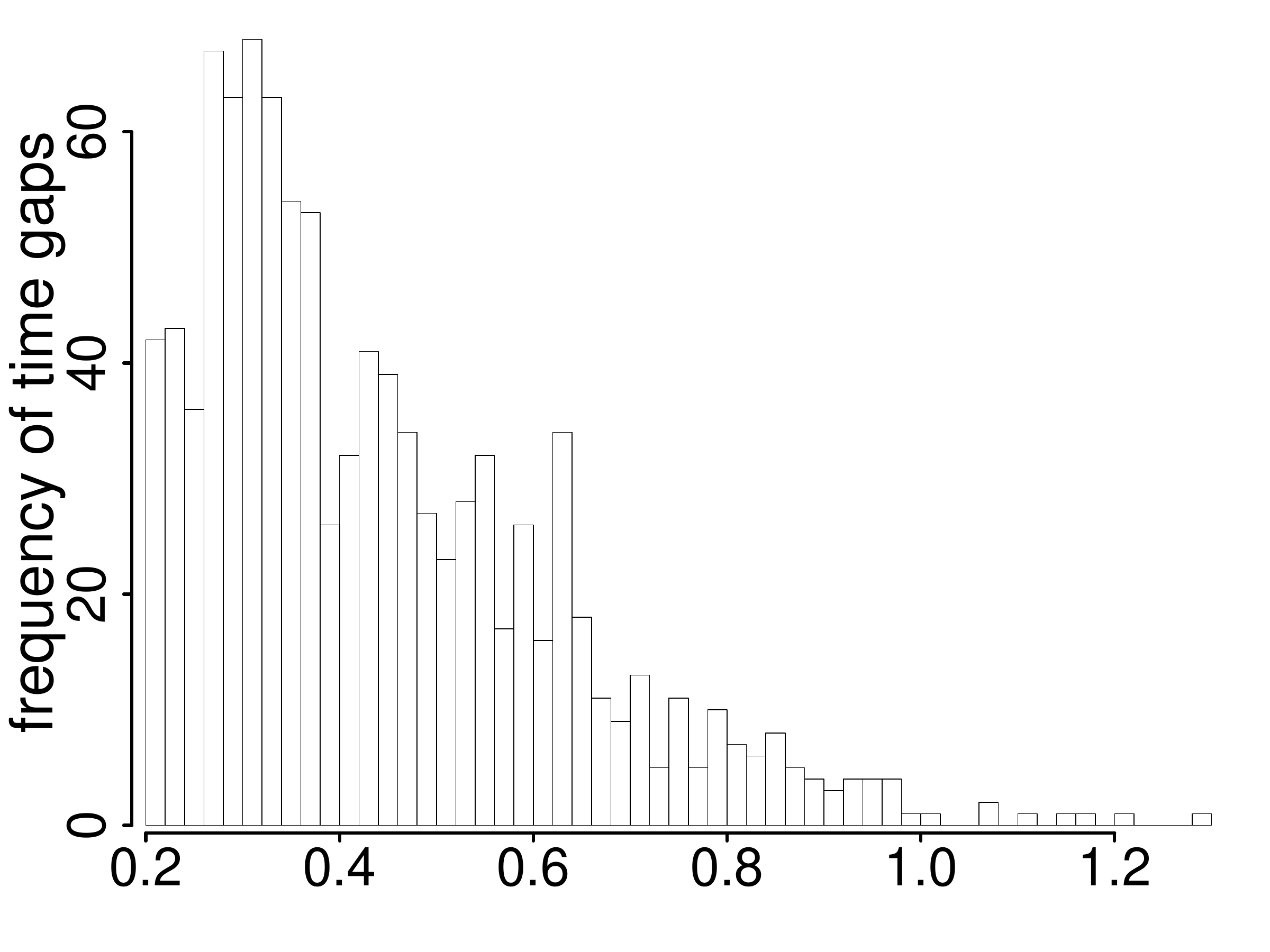}\\
\vspace{-6pt} (b) BVA TS \vspace{3pt}
\caption{Observed $\Y$, $\Delta \mathbf{t}$, and the histogram of $\Delta \mathbf{t}$} \label{fig:usdcad1}\vspace{-18pt}
\end{figure}

We apply the proposed COMS-GARCH and the reSAVE algorithm in two real data sets. The first is the exchange rate  between the US dollar and Canadian dollar collected from July to December 2008 and  can be downloaded from \url{https://www.histdata.com}.  The second is blood volume amplitude (BVA) TS data and can be downloaded from \url{https://archive.ics.uci.edu/ml/datasets/PPG-DaLiA#}. 
In the exchange rate application,  the volatility process of interest is the daily volatility of the log-return. 
To keep the data at a manageable size, we  took every $90$-th observation and performed a log transformation on the exchange rate. The final TS $\mathbf{G}$ we worked with contains $n=1501$ times points from July to December.  Figure \ref{fig:usdcad1}(a) depicts  the log-return TS $\Y$ (obtained by differencing $\mathbf{G}$), time gap $\Delta \mathbf{t}$ in days, and the histogram $\Delta \mathbf{t}$. The TS $\Y$ exhibits a change in volatility somewhere between September and October, an indication of multiple volatility states. The histogram suggests that $\Delta \mathbf{t}$ has a median around 0.07 days, but can be as short as just a couple of hours (the majority) and as long as $\ge2$ days due to the weekend and holiday effect.  In the BVA application, we extracted the measurements from the photoplethysmograph of the blood volume pulse (64 Hz; i.e., 64 times per second) by taking the valley and peak pulse values in each cycle and scaling them by $0.01$. 
We then took the first $1,000$ BVA  measurements from a random patient 
as the input TS $\Y$.  Figure \ref{fig:usdcad1}(b) depicts the TS $\Y$, time gap $\Delta \mathbf{t}$ and its histogram in the BVA TS.  The plot of TS $\Y$ suggests different volatilities, which can occur when a subject is in different physical conditions or experiences different emotional episodes.  In summary, Figure \ref{fig:usdcad1} suggests that the time points are irregularly spaced, and the TS' have multiple regimens in both cases, justifying the employment of the COMS-GARCH process to analyze such data.  

\begin{figure}[!htb]
\centering
(a) exchange rate TS \hspace{1.85in} (b) BVA TS\\
\includegraphics[width=0.54\linewidth]{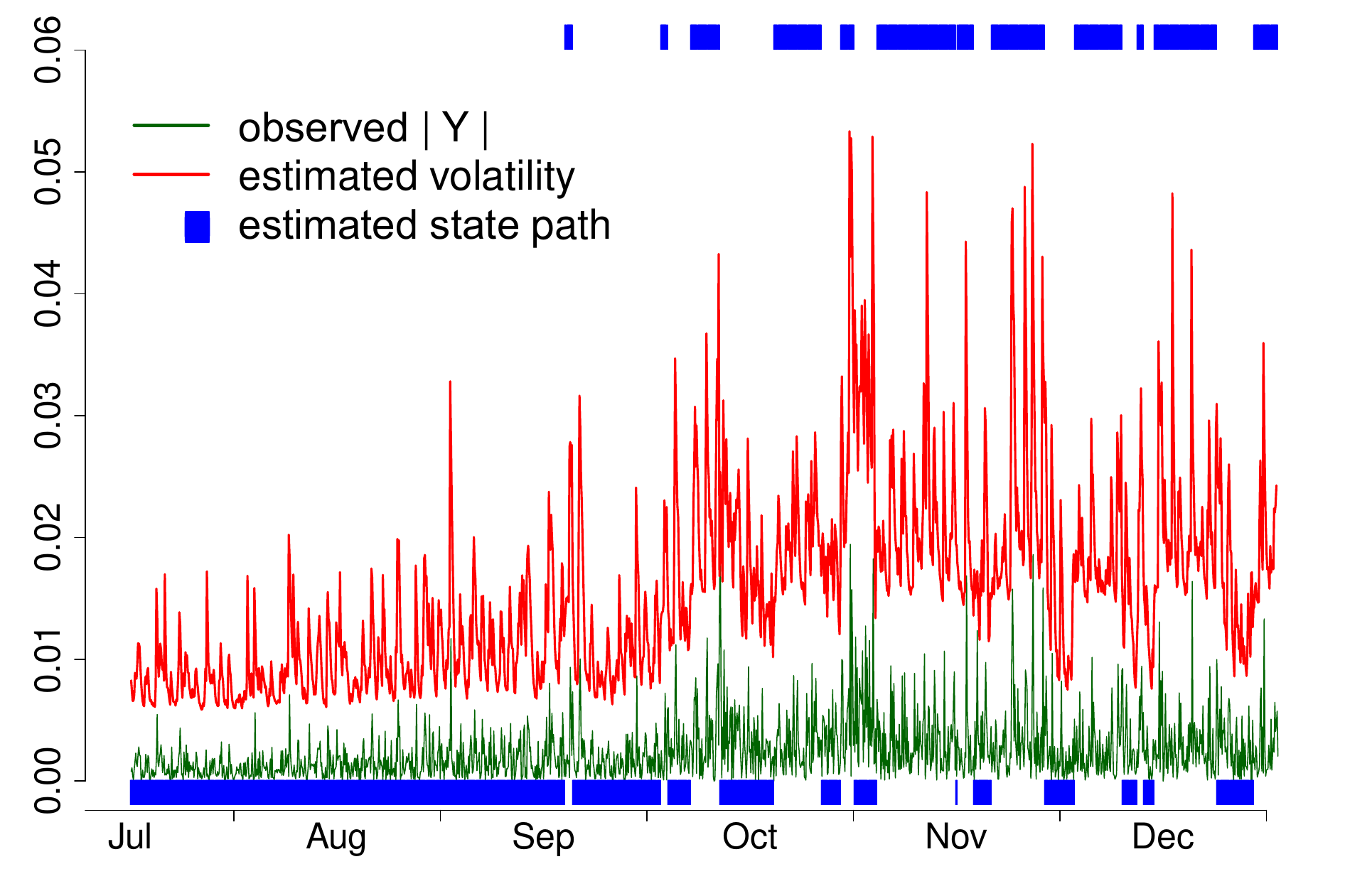}
\includegraphics[width=0.45\linewidth]{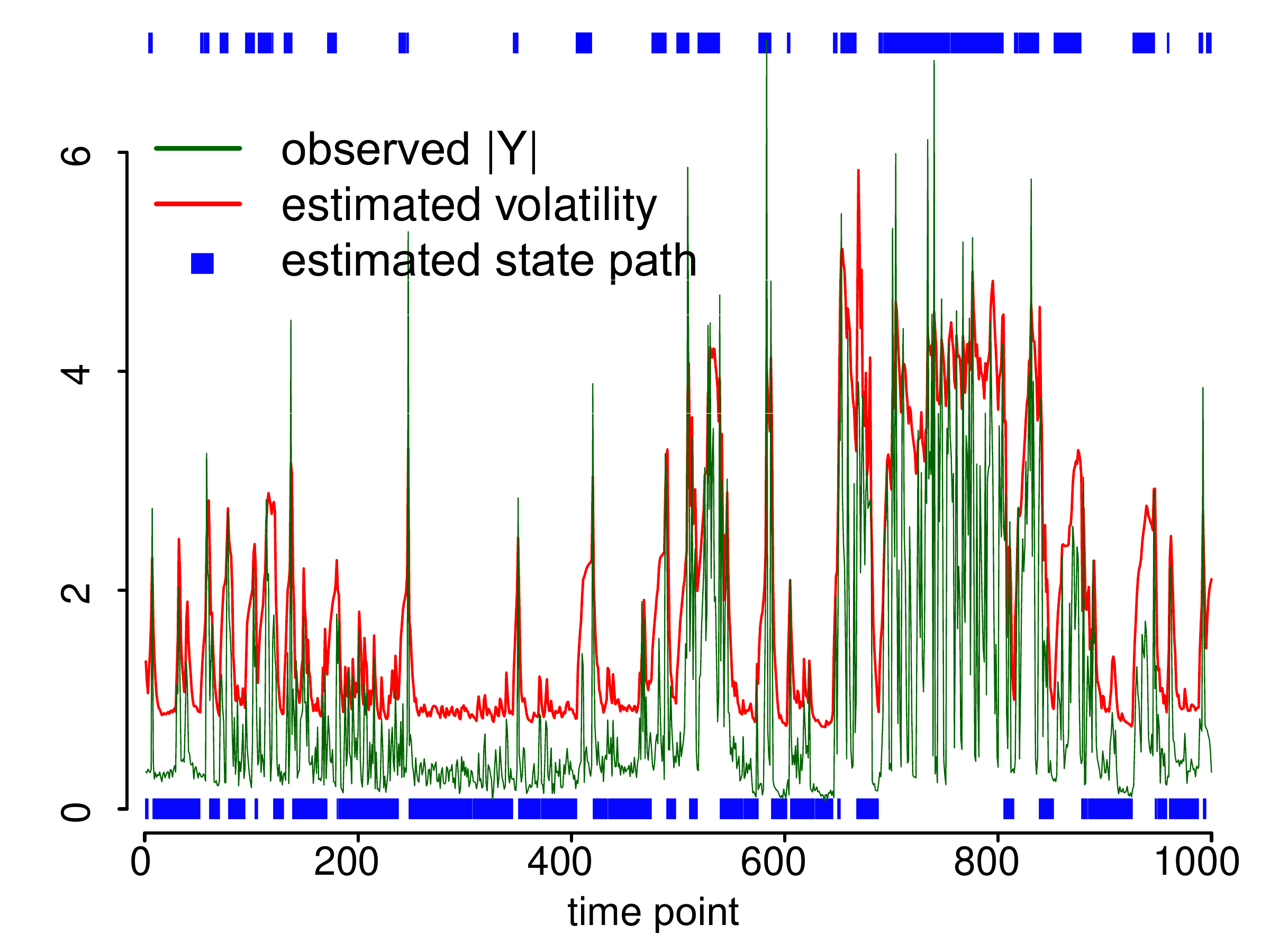}
\caption{Volatility and state estimation via the two-state COMS-GARCH and the reSAVE procedure}
\label{fig:usdcad2}\vspace{-9pt}
\end{figure}

The  $\Y$ plot in the exchange rate application suggests that there are two volatility states. We thus adopted a 2-state COMS-GARCH process to estimate the state path and volatilities via the reSAVE algorithm with $p=0.02, b=20,m=6$ and $N=1,000$ iterations. The  $\Y$ plot in the BVA application suggests that there are two volatility states. We adopted a 2-state COMS-GARCH process  to estimate the state path and volatilities via the reSAVE algorithm with $p=0.02, b=20,m=6$ and $N=1,500$ iterations.  The results from both applications are provided in Figure \ref{fig:usdcad2}. 
The estimated states and volatilities in general reflect well the two expected volatility states in both applications. For the exchange rate application, the estimated volatilities can be used to measure how far the exchange rate moves away from its mean value and help define risks and assess the volatility of the market when making investment decisions. For the BVA application, the estimated volatility and the state path can be used for evaluating how quickly a patient returns to the baseline (low) volatility state after spending a certain amount of time in the high-volatility state. A delay in returning hints potential cardiac risks in the patient. Compared to the simple methods such as the moving average \citep{bva}, the state and volatility predictions via the COMS-GARCH model are more interpretable as it assigns a definite state to each time point and does not have time-lagging issues when the TS switches regime. In both applications, the estimated state paths and volatilities can also be used as inputs for further analysis. 

\section{Discussion}\label{sec:discussion}
We propose the COMS-GARCH process for handling irregularly spaced TS data with multiple volatility states. We also introduce the reSAVE procedure with the Bernoulli NI for obtaining the MAP estimates for model parameters, state path, and volatilities. The computational efficiency and inferential robustness of the  reSAVE procedure are established and illustrated theoretically or empirically. 

As discussed briefly when presenting the simulations results, there is a lack of in-depth theoretical investigation on the asymptotic properties of the MLE and MAP estimation, such as consistency, based on the pseudo-likelihood or quasi-likelihood as $n\rightarrow\infty$ and $T\rightarrow\infty$ for the recently developed MS-GARCH and CO-GARCH processes; and the existence of bias in the estimation  of *-GARCH parameters is well acknowledged. This also creates opportunities for us to develop more accurate inferential procedures for *-GARCH models (COMS-GARCH included), coupled with investigation on their theoretical properties.

We conjecture that the reSAVE procedure is  applicable not only to the COMS-GARCH and CO-GARCH processes but also to other solvable CO-*-GARCH processes. For example, it will make an interesting future topic to develop the COMS-Exponential-GARCH and COMS-Integrated-GARCH processes and examine the  performance of the reSAVE procedure in these settings.  We also expect that the reSAVE procedure can be used in the COMS-ARMA process for trend estimation, yielding some types of weighted $l_2$ regularization on the ARMA parameters. More work will be needed to prove the theoretical conjectures formally.

\vspace{12pt}
\setstretch{1.0}
\bibliographystyle{apalike}
\bibliography{COMSGARCH}

\setstretch{1.1}
\section*{Appendix}
\appendix
\numberwithin{equation}{section}

{\large\textbf{Proof of Proposition \ref{prop:perturb1}}}

The conditional distribution of  $Y_i$ given $\Y_{i-1}, S,\Theta, \eta$ in the original TS is $N(0,\rho^2_i)$ (Eqn \ref{eqn:pseudo}), and that of $Y'_i$ from given $\Y'_{i-1}, S,\Theta, \eta$ in the perturbed TS is $N(0,\varrho^2_i)$, where $\Y'_{i-1}=(Y'_1,\ldots, Y'_{i-1})$, and $\varrho^2_i=\rho_i^2+\varepsilon^2$ for $i=1,\ldots,n$. Similarly,   $f(Y_{i+1}|\Y_i,S,\Theta, \eta)=N(0,\rho^2_{i+1})$ and $f(Y'_{i+1}|\Y'_i,S,\Theta, \eta)= N(0,\varrho^2_{i+1})$. WLOG, assume $\tilde{Y}_j=Y_i+Y_{i+1}$ after Bernoulli NI in the original TS and the corresponding perturbed version is $\tilde{Y}'_j=Y'_i+Y'_{i+1}$. 

When analyzing the perturbed TS and original perturbed TS data  after the NI via the same COMS-GARCH process, the contribution of observation $\tilde{Y}'_j$  and $\tilde{Y}_j$ to the overall negative log-pseudo-likelihood value given $S,\Theta,\bs\eta$ is 
\begin{align*}
&\log(L(\tilde{S},\Theta,\bs\eta,\tilde{Y}_j=Y_i+Y_{i+1}))= \frac{1}{2}\log(\tilde{\rho}_j^2)+\frac{(Y_i+Y_{i+1})^2}{2\tilde{\rho}_j^2}+\mbox{const.}\\
&\log(L(\tilde{S},\Theta,\bs\eta,\tilde{Y}'_j=Y'_i+Y'_{i+1}))= \frac{1}{2}\log(\tilde{\rho}_{j}^2)+\frac{(Y'_i+Y'_{i+1})^2}{2\tilde{\rho}_{j}^2}+\mbox{const.},
\end{align*}
respectively, and their expectations are 
\begin{align}
\!\!&\E\left(\log(L(S,\Theta,\bs\eta,\tilde{Y}_j\!=\!Y_i+Y_{i+1}))\right)\!=\! \frac{1}{2}\log(\tilde{\rho}_j^2)\!+\!\E\!\left(\!\frac{(Y_i+Y_{i+1})^2}{2\tilde{\rho}_j^2}\right)= \frac{1}{2}\log(\tilde{\rho}_j^2)\!+\!\frac{1}{2}\label{eqn:Eo}\\
\!\!&\E\left(\log(L(S,\Theta,\bs\eta,\tilde{Y}'_j\!=\!Y'_i+Y'_{i+1})\right)\!=\!\frac{1}{2}\log(\tilde{\rho}_j^2)+\!\E\!\left(\frac{(Y'_i+Y'_{i+1})^2}{2\tilde{\rho}_j^2}\right)\!=\! \frac{1}{2}\log(\tilde{\rho}_j^2)\!+\!\frac{1}{2}\!+\!\frac{\varepsilon^2}{\tilde{\rho}_j^2}.\label{eqn:Ep}
\end{align}
The difference between Eqns (\ref{eqn:Ep}) and (\ref{eqn:Eo}) 
\begin{equation}
\varepsilon^2/(\tilde{\rho}_j^2)=\varepsilon^2/(\rho_i^2+\rho_{i+1}^2)\label{eqn:dNI}
\end{equation}
Without Bernoulli NI, the contribution of $Y_{i+1}$ and $Y_i$ to the overall negative log-pseudo-likelihood on $S,\Theta,\bs\eta$, given the original perturbed data is, respectively
$$\log\left(L(S,\Theta,\bs\eta,Y_{i+1})\right)
=\frac{1}{2}\log(\rho_{i+1}^2)+\frac{Y^2_{i+1}}{2\rho_{i+1}^2} \mbox{ and } 
\log\left(L(S,\Theta,\bs\eta,Y_i)\right)
=\frac{1}{2}\log(\rho_i^2)+\frac{Y^2_i}{2\rho_i^2}.$$
And their expectations are 
\begin{align}
\E\left(\log\left(L(S,\Theta,\bs\eta,Y_{i+1})\right)\right)
=&\frac{1}{2}\log(\rho_{i+1}^2) +\E\left(\frac{Y^2_{i+1}}{2\rho_{i+1}^2}\right)
=\frac{1}{2}\log(\rho_{i+1}^2) +\frac{1}{2},\label{eqn:o1}\\
\E\left(\log\left(L(S,\Theta,\bs\eta,Y_i)\right)\right)
=&\frac{1}{2}\log(\rho_i^2) +\E\left(\frac{Y^2_i}{2\rho_i^2}\right)
=\frac{1}{2}\log(\rho_i^2) +\frac{1}{2},\label{eqn:o2}
\end{align}
respectively. Similarly, in the perturbed TS $\Y'$, 
\begin{align}
&\E\left(\log\left(L(S,\Theta,\bs\eta,Y'_{i+1})\right)\right)
=\frac{1}{2}\log(\rho_{i+1}^2) +\E\left(\frac{Y^{'2}_{i+1}}{2\rho_{i+1}^2}\right)=\frac{1}{2}\log(\rho_{i+1}^2)+\frac{1}{2}+\frac{\varepsilon^2}{2{\rho}_{i+1}^2}\label{eqn:p1}\\
&\E\left(\log\left(L(S,\Theta,\bs\eta,Y'_i)\right)\right)
=\frac{1}{2}\log(\rho_i^2)+\E\left(\frac{Y^{'2}_i}{2\rho_i^2}\right)=\frac{1}{2}\log(\rho_i^2) +\frac{1}{2}+\frac{\varepsilon^2}{2{\rho}_i^2}\label{eqn:p2}.
\end{align}
The difference in the sum of the negative conditional log-pseudo-likelihood given $Y'_i$ and $Y'_{i+1}$ vs. that given $Y_i$ and $Y_{i+1}$, that is, Eqns [(\ref{eqn:p1})+(\ref{eqn:p2})]-[(\ref{eqn:o1})+(\ref{eqn:o2})], is 
\begin{equation}
\varepsilon^2/(2{\rho}_i^2)+\varepsilon^2/(2{\rho}_{i+1}^2)\label{eqn:d}
\end{equation}
Finally, the ratio  between Eqns (\ref{eqn:dNI}) and (\ref{eqn:d}) is $\frac{(2\rho_i)^{-2}+(2\rho_{i+1})^{-2}}{\left(\rho_i^2+\rho_{i+1}^2\right)^{-1}}= \frac{2\rho_i^2\rho_{i+1}^2}{\left(\rho_i^2+\rho_{i+1}^2\right)^2}<1.$; in other words, the difference in the expected loss function between the original TS vs. that subject to random perturbation with Bernoulli NI is smaller that without Bernoulli NI. 

In general, when applying the Bernoulli NI during the reSAVE procedure, a time point is dropped from the TS randomly with a probability $p$. Say $r$ observations are dropped between $G_{i-1}$ and $G_{i+r}$, then $\tilde{Y}_{j}=  Y_{i+r}+\cdots+Y_i$.  Eqns (\ref{eqn:d}) and (\ref{eqn:dNI})  now become
\begin{align}\label{eqn:generalr}
\textstyle \sum_{k=0}^r\varepsilon^2/(2{\rho}_k^2) \mbox{ and }\varepsilon^2/(2\sum_{k=0}^r{\rho}_{i+k}^2), 
\end{align}
respectively and their ratio is
\begin{align*}
&\sum_{k=0}^r\frac{\varepsilon^2}{2{\rho}_{i+k}^2}\bigg/\frac{r\varepsilon^2}{2\sum_{k=0}^r{\rho}_{i+k}^2}=
r^{-1}\sum_{k=0}^r{\rho}_{i+k}^{-2}\sum_{k=0}^r{\rho}_{i+k}^2=r^{-1}\sum_{k=0}^r\frac{\sum_{k=0}^r{\rho}_{i+k}^2}{{\rho}_{i+k}^2}>1
\end{align*}
for any general $r\ge1$ in each sub-TS  that involves a dropped observation. Putting  all the sub-TS' together, the overall difference in the expected negative log-likelihood function with vs without perturbation, after Bernoulli NI, is smaller than that without Bernoulli NI. 

In the framework of Bayesian modelling,  the  objective function becomes the log-posterior distribution, that is the sum of the log-likelihood function and the log-prior. Since the prior is the same with vs. without the external perturbation,  it is cancelled out when the difference between two log-posterior distributions is taken, and we arrive at the  same equations as Eqns (\ref{eqn:dNI}), (\ref{eqn:d}), and (\ref{eqn:generalr}); and the conclusion also holds in the Bayesian framework.
\qed

\vspace{12pt}

\clearpage
\setcounter{page}{1}
\setcounter{figure}{0}
\setcounter{table}{0}
\setcounter{section}{0}
\renewcommand{\thetable}{S\arabic{table}}
\renewcommand{\thefigure}{S\arabic{figure}}
\renewcommand{\thesection}{S\arabic{section}}    

\centering\textbf{\LARGE{Supplementary Materials}}\\

\justify
\section{EM and MC-EM Algorithms for Parameter Estimation in COMS-GARCH process}\label{sec:em}
The expectation step in iteration $l$ of the EM algorithm comprises the calculation of \vspace{-12pt}
\begin{align}
&\E_S\left(L(\Theta,S|\bs{\Delta t},\Y, \Theta^{(l-1)},\bs\eta^{(l-1))}\right)\notag\\
=&\textstyle\sum_{S\in\mathcal{S}}L(\Theta|\bs{\Delta t},\Y,S) \times \Pr(S|\bs{\Delta t},\Y,\Theta^{(l-1)},\bs\eta^{(l-1)})\notag\\
\propto&\textstyle\sum_{S\in\mathcal{S}}\left(  \prod_{i=1}^{n}\rho_i^{-1}\exp\left( -Y_i^2/(2\rho_i^2) \right)\right) \cdot\Pr(S|\Y,\Theta^{(l-1)},\bs\eta^{(l-1)}), \mbox{ and}\label{eqn:Etheta}\\
&E_S\left(L(\bs\eta,S|\bs{\Delta t},\Y, \Theta^{(l-1)},\bs\eta^{(l-1)}\right)\notag\\
=&\textstyle \sum_{S\in\mathcal{S}} \left\{\prod_{i=2}^{n}\left[\left(1\!-\!\exp(-\eta_{s_i,s_{{i-1}}}\Delta t_i)\right)^{\1(s_i\neq s_{{i-1}})}\left(2\!-\!\nu\!+\!\sum_{v\neq s_{{i-1}}}\exp(-\eta_{v,s_{{i-1}}}\Delta t_i)\right)^{\1(s_i= s_{{i-1}})}\right]\right.\notag\\
&\qquad\qquad\left.\times\textstyle\Pr(S|\bs{\Delta t},\Y,\Theta^{(l-1)},\bs\eta^{(l-1)})\right\}\label{eqn:Eeta}, 
\end{align}
where $\bs{\Delta t}=(\Delta t_1,\ldots,\Delta t_n),\Y=(Y_1,\ldots,Y_n)$, and $\mathcal{S}$ is the set of all possible paths. The M-step maximizes the expectations in Eqns (\ref{eqn:Etheta}) and (\ref{eqn:Eeta}) to obtain MLEs $\Theta^{(l)}$ and $\bs\eta^{(l)}$. 

The classical EM estimation procedure can be computationally intensive or even unfeasible  when $n$ is large. In the E-step, one would sum over $\nu^n$ possible state paths in Eqns (\ref{eqn:Etheta}) and (\ref{eqn:Eeta}). For example, if $\nu=2$ and $n=100$, the number of possible state paths is $3.27\times 10^{150}$, an astronomical figure to deal with in practice. On top of that, $\sigma^2_i$ and  $\rho^2_i$ for a given path are calculated  recursively for $i=1,\ldots,n$ (Eqns (\ref{eqn:sigma2n}) and (\ref{eqn:approx})).  To circumvent the computational issue, the MC-EM algorithm can be used to obtain numerical approximations to the expected likelihood. Specifically, the expectations over $S\in\mathcal{S}$ in Eqns (\ref{eqn:Etheta}) and (\ref{eqn:Eeta}) are replaced by their respective averages over a set of path samples $S_j\!=\!(s_{1j},\ldots,s_{nj})$ for $j\!=\!1,\ldots,m$ in each iteration. Specifically, in the $l$-th iteration, Eqns (\ref{eqn:Etheta}) and (\ref{eqn:Eeta}) will be replaced by
\begin{align*}
&\textstyle m^{-1}\sum_{j=1}^m L(\Theta|\bs{\Delta t},\Y,S_j^{(l-1)}) \times \Pr(S_j^{(l-1)}|\bs{\Delta t},\Y,\Theta^{(l-1)},\bs\eta^{(l-1)})\\
&\textstyle m^{-1}\! \sum_{j=1}^m\! \left\{\!\prod_{i=2}^{n}\!\!\left[\!\left(\!1\!-\!\exp(-\eta_{s^{(l-1)}_{ij},s^{(l-1)}_{{i-1,j}}}\Delta t_i)\right)^{\!\1(s^{(l-1)}_{ij}\neq s^{(l-1)}_{{i-1,j}})}\times \right.\right.\\
&\qquad\qquad\textstyle\left.\left.\left(2\!-\!\nu\!+\!\sum_{v\neq s^{(l-1)}_{i,j}}\exp(-\eta_{v,s^{(l-1)}_{i-1,j}}\Delta t_i)\right)^{\!\1(s^{(l-1)}_{i,j}= s^{(l-1)}_{i-1,j)}}\right]\Pr(S_j^{(l-1)}|\bs{\Delta t},\Y,\Theta^{(l-1)},\bs\eta^{(l-1)})\right\}.
\end{align*}
The path samples $S_j$  for $j=1,\ldots,m$ are drawn from the conditional distribution of state $s_i$ for $i=1,\ldots,n$ given $\Y,\mathbf{\Delta t}, \Theta^{(l)},\bs\eta^{(l-1)}$, and the states at other time points $S^{(l-1)}_{-i}=\left\{s^{(l-1)}_1,\ldots, s^{(l-1)}_{i-1},s^{(l-1)}_{i+1},\ldots, s^{(l-1)}_n\right\}$. Specifically, 
\begin{align}
&s^{(l)}_i\sim f({s_i}|S^{(l-1)}_{-i},\Theta^{(l)},\bs\eta^{(l-1)}, \Y,\mathbf{\Delta t})
\propto\textstyle \xi_{s_i,s_{{i-1}}}\xi_{s_{i+1},s_i}\prod_{t=i}^{n}\rho_t^{-1}(s_t)\exp\!\left(\!-
Y_t^2/(2\rho_t^2) \right),\label{eqn:statei}\\ 
&\mbox{where }\xi_{s_i,s_{i-1}}\!=\!\begin{cases}
2\!-\!\nu\!+\!\sum_{k\ne  s_{i-1}}\!\!\exp(-\eta_{k,s_{i-1}}\Delta t_i) & \mbox{when }  s_i=s_{i-1}\\
1-\exp(-\eta_{s_i,s_{i-1}}\Delta t_i) & \mbox{when }s_i\neq s_{i-1}\\
\end{cases}\!;
\mbox{ similarly for }\xi_{s_{i+1},s_i}.\notag
\end{align}

\section{Additional Results From the Simulation Studies}
\begin{figure}[!htb]
\centering
\includegraphics[width=.75\textwidth]{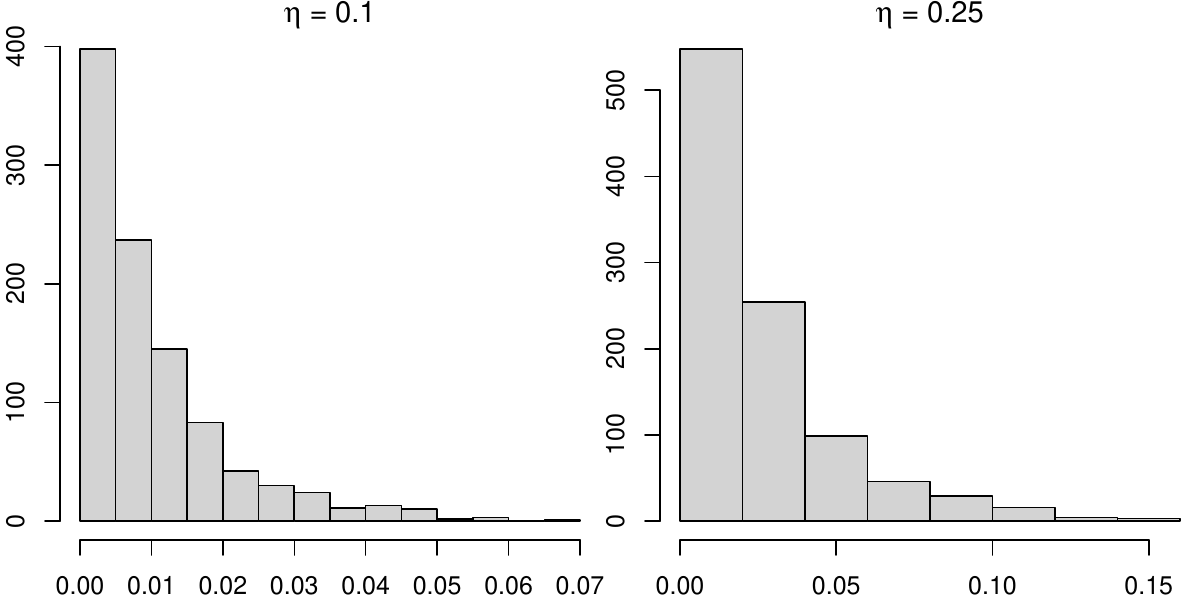}
\caption{Histogram of $\Pr(s_1|s_2)=\Pr(s_2|s_1)$ in Simulation study 2} \label{fig:transitPhist}
\end{figure}

\begin{figure}[!htb]
\hspace{1in}($m=1, p=0$) \raisebox{-0.5\height}{\includegraphics[trim=0.2cm 0 0.6cm 0, clip, width=0.45\textwidth]{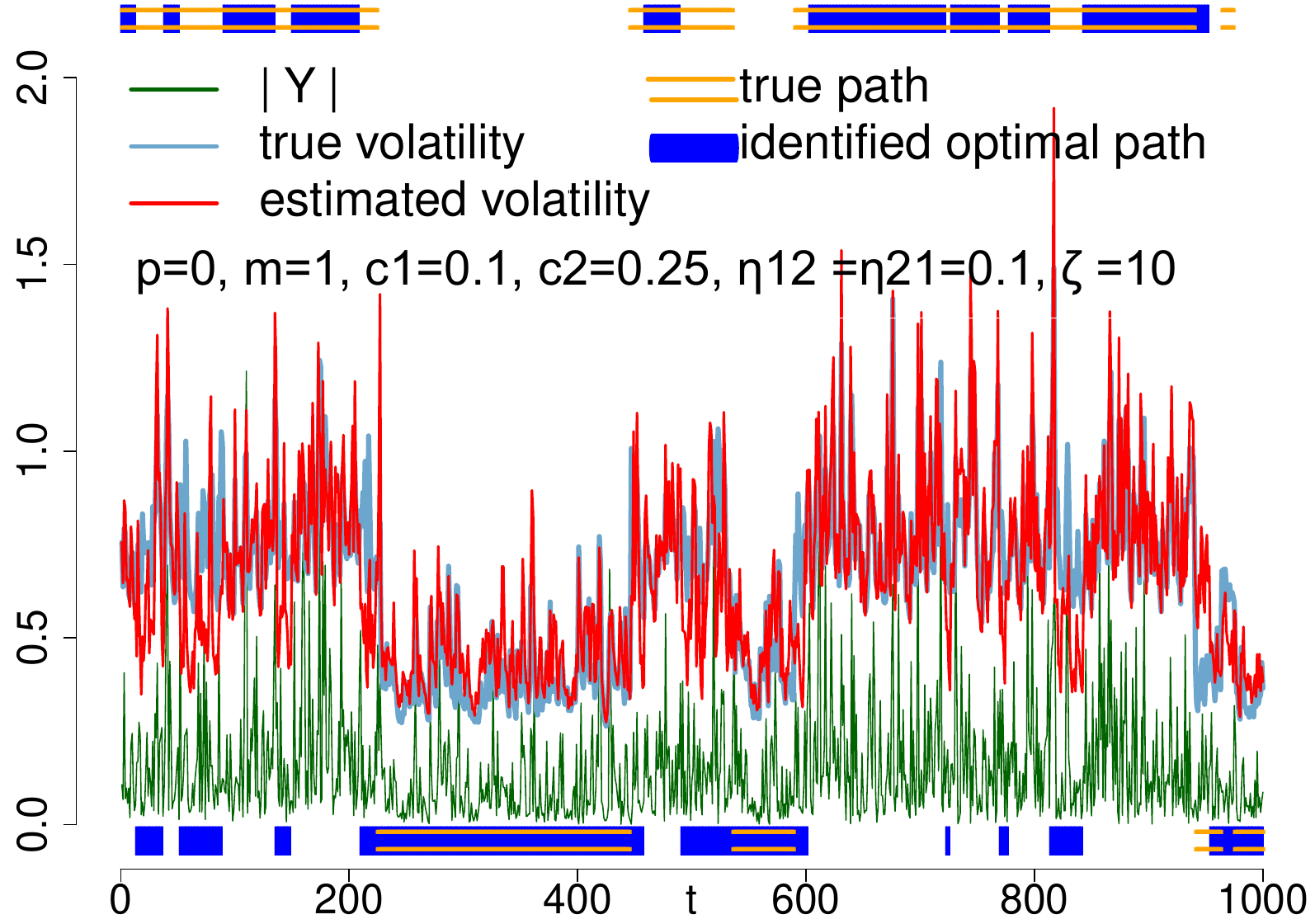}}\\

\hspace{1in} ($m=6, p=0$) \raisebox{-0.5\height}{\includegraphics[trim=0.2cm 0 0.6cm 0, clip, width=0.45\textwidth]{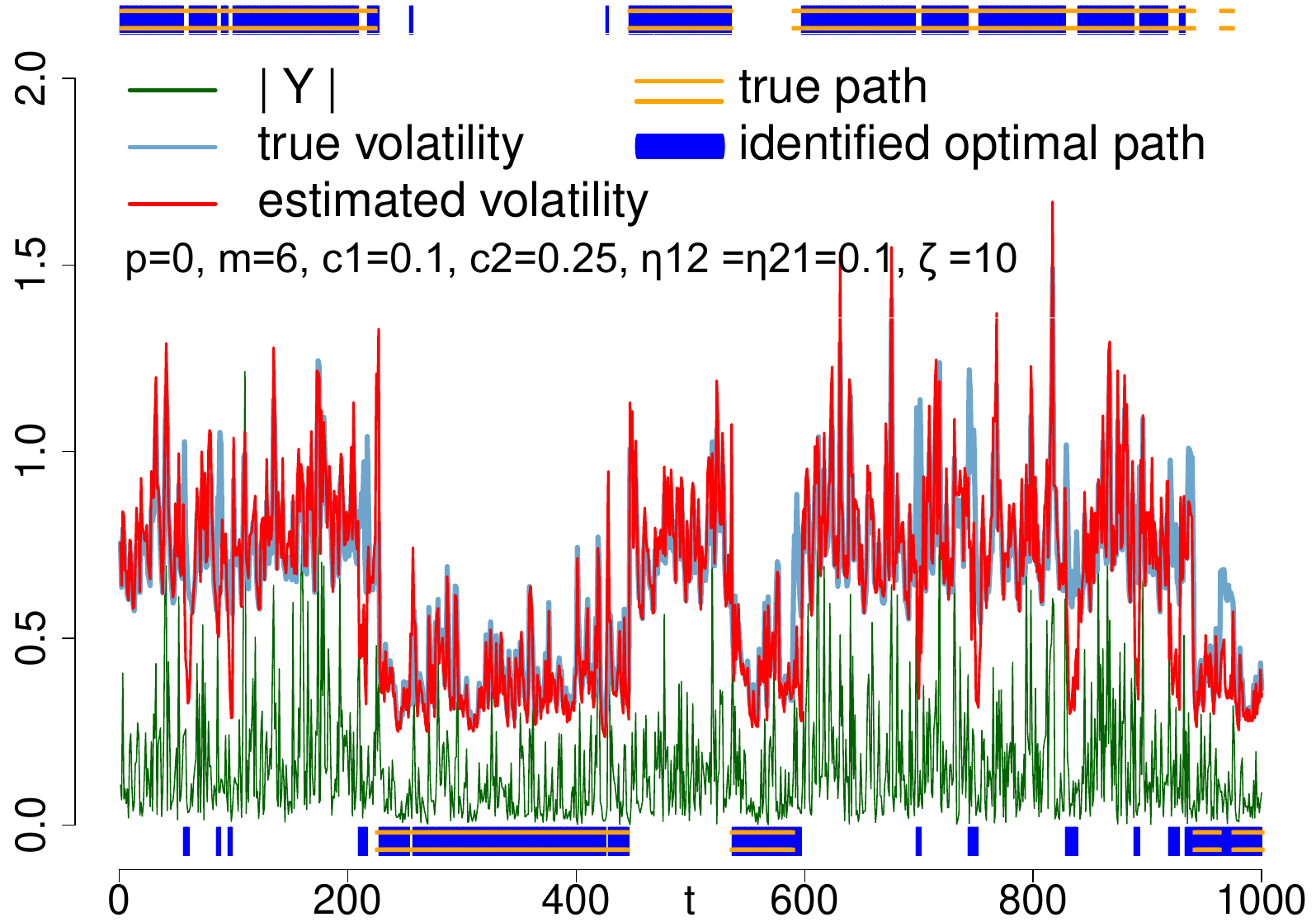}}\\

\hspace{1in}($m=6, p=0.02$) \raisebox{-0.5\height}{\includegraphics[trim=0.25cm 0 0.6cm 0, clip, width=0.45\textwidth]{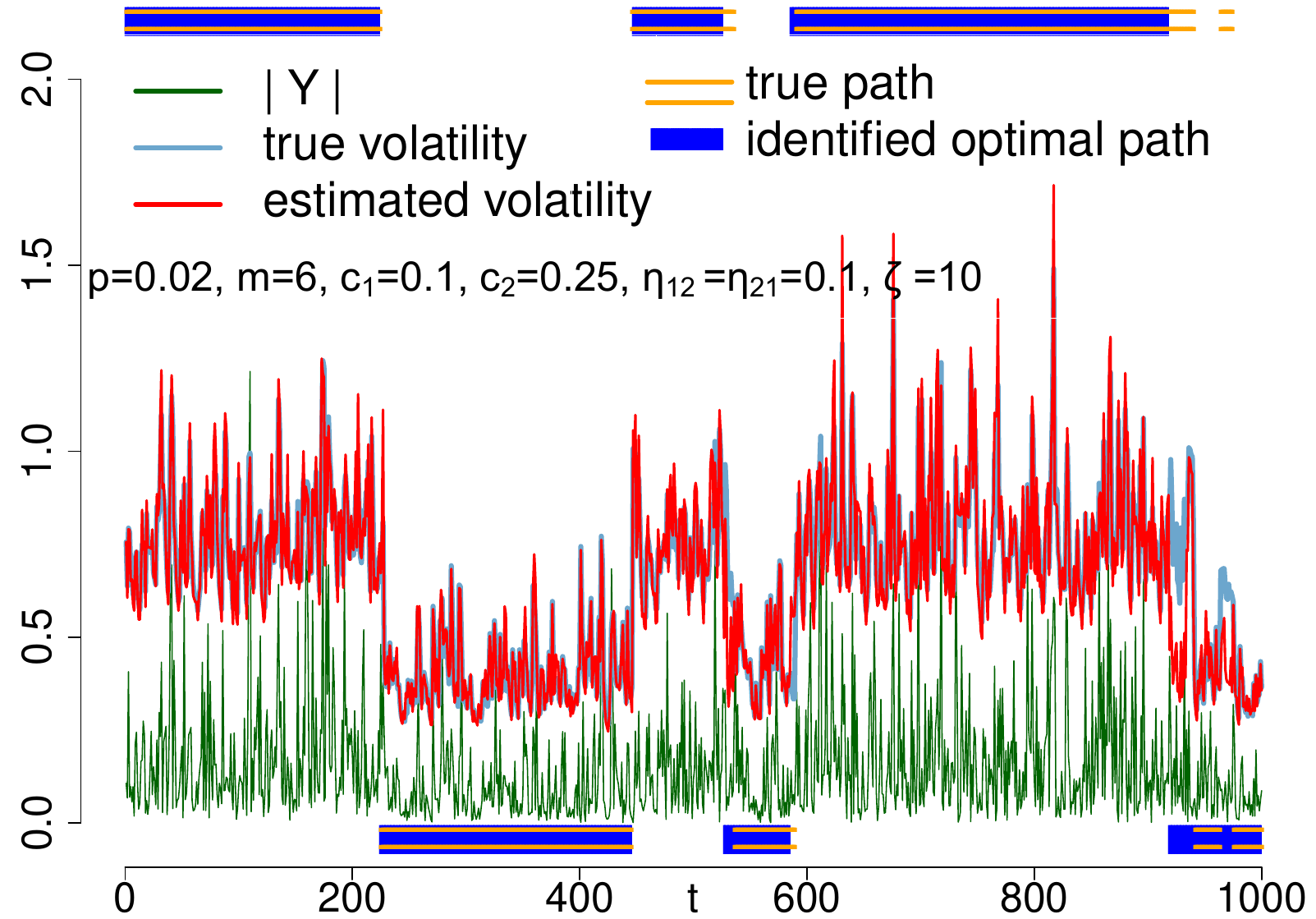}}
\caption{Estimated volatility and state path from one repetition in Simulation Study 2. The estimates overlap well with the true volatilities and state path except at a few time points}\label{fig:SIM2mtrace}\vspace{-6pt}
\end{figure}

\end{document}



\renewcommand{\baselinestretch}{2}

\markright{ \hbox{\footnotesize\rm Statistica Sinica: Supplement
}\hfill\\[-13pt]
\hbox{\footnotesize\rm
}\hfill }

\markboth{\hfill{\footnotesize\rm FIRSTNAME1 LASTNAME1 AND FIRSTNAME2 LASTNAME2} \hfill}
{\hfill {\footnotesize\rm FILL IN A SHORT RUNNING TITLE} \hfill}

\renewcommand{\thefootnote}{}
$\ $\par \fontsize{12}{14pt plus.8pt minus .6pt}\selectfont


 \centerline{\large\bf THE TITLE OF THE MANUSCRIPT}
\vspace{2pt}
 \centerline{\large\bf IF SECOND LINE IS NEEDED}
\vspace{2pt}
 \centerline{\large\bf IF THIRD LINE IS NEEDED}
\vspace{.25cm}
 \author{Author(s)}
\vspace{.4cm}
 \centerline{\it The affiliation(s) of authors}
\vspace{.55cm}
 \centerline{\bf Supplementary Material}
\vspace{.55cm}
\fontsize{9}{11.5pt plus.8pt minus .6pt}\selectfont
\noindent
CONTENT OF A BRIEF NOTE.........\\
CONTENT OF THE BRIEF NOTE.\\
CONTENT OF THE BRIEF NOTE.\\
\par

\setcounter{section}{0}
\setcounter{equation}{0}
\def\theequation{S\arabic{section}.\arabic{equation}}
\def\thesection{S\arabic{section}}

\fontsize{12}{14pt plus.8pt minus .6pt}\selectfont

\section{Title of section 1}

CONTENT OF THIS SECTION........\\
CONTENT OF THIS SECTION.\\
CONTENT OF THIS SECTION.\\
\par
\begin{equation}
\mbox {The 1st display equation of section 1.}
\end{equation}

\begin{equation}
\mbox {The 2nd display equation of section 1.}
\end{equation}

\section{Title of section 2}
\setcounter{equation}{0}

CONTENT OF THIS SECTION........\\
CONTENT OF THIS SECTION.\\
CONTENT OF THIS SECTION.\\
\par

\begin{equation}
\mbox {The 1st display equation of S.2.}
\end{equation}

\begin{equation}
\mbox {The 2nd display equation of S.2.}
\end{equation}

\newpage
\lhead[\footnotesize\thepage\fancyplain{}\leftmark]{}\rhead[]{\fancyplain{}\rightmark\footnotesize\thepage}

\section{Title of section 3}
\setcounter{equation}{0}

\section{Title of section 4}
\setcounter{equation}{0}